\documentclass[11pt, a4paper]{article}
\pdfoutput=1
\usepackage{jheppub,amsmath,amssymb,slashed,url,bm,textgreek,upgreek}
\usepackage{jheppub}  
\usepackage{tikz,lipsum,lmodern}
\usepackage[most]{tcolorbox}
\usepackage{amssymb} 
\usepackage{amsmath}
\usepackage{mathtools}
\usepackage{amsfonts}    
\usepackage{dsfont}
\usepackage{pdfpages}
\usepackage{verbatim}
\usepackage{tensor}
\usepackage{mathrsfs}
\usepackage{textgreek} 
\usepackage[mathscr]{euscript}
\usepackage[normalem]{ulem}
\usepackage{tikz}
\usepackage{makecell}
\usepackage[verbose]{placeins}
\usepackage{subcaption}
\usetikzlibrary{3d, arrows.meta, decorations.pathreplacing, decorations.markings,calc,shapes.misc,decorations.pathmorphing,patterns.meta, math}
\usepackage{pgfplots}
\pgfplotsset{compat=1.18}

\newcommand{\beq}{\begin{equation}}
\newcommand{\eeq}{\end{equation}}
\newcommand{\nn}{\nonumber\\} 
\newcommand{\bea}{\begin{eqnarray}}
\newcommand{\ea}{\end{eqnarray}}
\newcommand{\barr}{\begin{array}}
\newcommand{\earr}{\end{array}}

\def\be{\begin{equation}}
\def\ee{\end{equation}}
\def\ba#1\ea{\begin{align}#1\end{align}}
\def\bg#1\eg{\begin{gather}#1\end{gather}}
\def\bm#1\em{\begin{multline}#1\end{multline}}
\def\bmd#1\emd{\begin{multlined}#1\end{multlined}}

\def\d{{\rm d}}
\def\i{{\rm i}}
 \def\L{\ell}

\newcommand\U{\text{U}}
\newcommand\SL{\text{SL}}
\newcommand\SU{\text{SU}}

\def\vp{\chi}

\def\la{\label}

\def\wg{\wedge}

\def\no{\nonumber}

\def\({\left(}
\def\){\right)}
\def\[{\left[}
\def\]{\right]}
\def\<{\langle}
\def\>{\rangle}

\def\bea{\begin{eqnarray}}
\def\eea{\end{eqnarray}}

\newcommand{\tr}{\operatorname{tr}}

\newcommand{\zb}{{\bar z}}

\def\nn{\nonumber}

\def\Re{{\rm Re}}

\begin{document}

\global\long\def\aad{(a\tilde{a}+a^{\dagger}\tilde{a}^{\dagger})}%

\global\long\def\ad{{\rm ad}}%

\global\long\def\bij{\langle ij\rangle}%

\global\long\def\df{\coloneqq}%

\global\long\def\bs{b_{\alpha}^{*}}%

\global\long\def\bra{\langle}%

\global\long\def\dd{{\rm d}}%

\global\long\def\dg{{\rm {\rm \dot{\gamma}}}}%

\global\long\def\ddt{\frac{{\rm d^{2}}}{{\rm d}t^{2}}}%

\global\long\def\ddg{\nabla_{\dot{\gamma}}}%

\global\long\def\del{\mathcal{\delta}}%

\global\long\def\Del{\Delta}%

\global\long\def\dtau{\frac{\dd^{2}}{\dd\tau^{2}}}%

\global\long\def\ul{U(\Lambda)}%

\global\long\def\udl{U^{\dagger}(\Lambda)}%

\global\long\def\dl{D(\Lambda)}%

\global\long\def\da{\dagger}%

\global\long\def\id{{\rm id}}%

\global\long\def\ml{\mathcal{L}}%

\global\long\def\mm{\mathcal{\mathcal{M}}}%

\global\long\def\mf{\mathcal{\mathcal{F}}}%

\global\long\def\ket{\rangle}%

\global\long\def\kpp{k^{\prime}}%

\global\long\def\lr{\leftrightarrow}%

\global\long\def\lf{\leftrightarrow}%

\global\long\def\ma{\mathcal{A}}%

\global\long\def\mb{\mathcal{B}}%

\global\long\def\md{\mathcal{D}}%

\global\long\def\mbr{\mathbb{R}}%

\global\long\def\mbz{\mathbb{Z}}%

\global\long\def\mh{\mathcal{\mathcal{H}}}%

\global\long\def\mi{\mathcal{\mathcal{I}}}%

\global\long\def\ms{\mathcal{\mathcal{\mathcal{S}}}}%

\global\long\def\mg{\mathcal{\mathcal{G}}}%

\global\long\def\mfa{\mathcal{\mathfrak{a}}}%

\global\long\def\mfb{\mathcal{\mathfrak{b}}}%

\global\long\def\mfb{\mathcal{\mathfrak{b}}}%

\global\long\def\mfg{\mathcal{\mathfrak{g}}}%

\global\long\def\mj{\mathcal{\mathcal{J}}}%

\global\long\def\mk{\mathcal{K}}%

\global\long\def\mmp{\mathcal{\mathcal{P}}}%

\global\long\def\mn{\mathcal{\mathcal{\mathcal{N}}}}%

\global\long\def\mq{\mathcal{\mathcal{Q}}}%

\global\long\def\mo{\mathcal{O}}%

\global\long\def\qq{\mathcal{\mathcal{\mathcal{\quad}}}}%

\global\long\def\ww{\wedge}%

\global\long\def\ka{\kappa}%

\global\long\def\nn{\nabla}%

\global\long\def\nb{\overline{\nabla}}%

\global\long\def\pathint{\langle x_{f},t_{f}|x_{i},t_{i}\rangle}%

\global\long\def\ppp{p^{\prime}}%

\global\long\def\qpp{q^{\prime}}%

\global\long\def\we{\wedge}%

\global\long\def\pp{\prime}%

\global\long\def\sq{\square}%

\global\long\def\vp{\varphi}%

\global\long\def\ti{\widetilde{}}%

\global\long\def\wg{\widetilde{g}}%

\global\long\def\te{\theta}%

\global\long\def\tr{{\rm Tr}}%

\global\long\def\ta{{\rm \widetilde{\alpha}}}%

\global\long\def\sh{{\rm {\rm sh}}}%

\global\long\def\ch{{\rm ch}}%

\global\long\def\Si{{\rm {\rm \Sigma}}}%

\global\long\def\sch{{\rm {\rm Sch}}}%

\global\long\def\vol{{\rm {\rm {\rm Vol}}}}%

\global\long\def\reg{{\rm {\rm reg}}}%

\global\long\def\zb{{\rm {\rm |0(\beta)\ket}}}%

\title{The wavefunction of a quantum $S^1 \times S^2$ universe
\vspace{-0.8cm}}

\author{Gustavo J. Turiaci and Chih-Hung Wu}
\affiliation{Department of Physics, University of Washington, Seattle, WA 98195, USA}
\emailAdd{turiaci@uw.edu, chwu29@uw.edu}

\abstract{We study quantum gravity corrections to the no-boundary wavefunction describing a universe with spatial topology $S^1\times S^2$. It has been suggested that quantum effects become increasingly important when the size of the circle is large relative to the sphere. In this paper, we confirm this claim by an explicit four-dimensional one-loop calculation of the gravitational path integral preparing such a state. In the process, we clarify some aspects of the gravitational path integral on complex spacetimes. These quantum corrections play a crucial role in ensuring that the norm of the wavefunction is naturally expressed in terms of a path integral over $S^2 \times S^2$ at the classical level. We extend some of the analysis to more general spatial topologies, as well as to the inclusion of fermions.}

\maketitle

\newpage

\section{Introduction}

Recent developments in quantum aspects of black holes have pointed towards the important role of the gravitational path integral in quantum gravity. Originally proposed by Gibbons and Hawking \cite{Gibbons:1976ue} as a way to reproduce some classical features of black hole thermodynamics, it has been recently used for entanglement entropy computations in the context of AdS/CFT \cite{Lewkowycz:2013nqa}, for the calculation of the fine-grained entropy of Hawking radiation \cite{Almheiri:2019qdq,Penington:2019kki}, to derive quantum chaotic features of the black hole spectrum such as level repulsion \cite{Saad:2018bqo,Saad:2019lba}, among others.  

\smallskip

A particular application of the gravitational path integral, relevant to the present paper, is to the spectrum of near-extremal charged black holes \cite{Ghosh:2019rcj,Iliesiu:2020qvm,Heydeman:2020hhw,Castro:2021wzn,David:2021qaa, Boruch:2022tno, Iliesiu:2022onk,Turiaci:2023wrh,Aggarwal:2023peg} as well as rotating ones \cite{Castro:2021csm, Rakic:2023vhv, Kapec:2023ruw, Maulik:2024dwq,Arnaudo:2024bbd}. Quantum corrections, that as far as we know can only be captured by the gravitational path integral formalism, resolve certain puzzles that arise in the limit where the temperature of charged black holes vanishes \cite{Preskill:1991tb,Maldacena:1998uz,Page:2000dk}.

\smallskip 

Although the last decade has seen tremendous progress on the black hole front, solid progress in applying the gravitational path integral to cosmology has certainly been a more difficult task. This program started with the Hartle and Hawking proposal of the no-boundary wavefunction of the universe \cite{Hartle:1983ai}. Conceptually, this idea parallels the same development that worked so well for black holes, but the results are often difficult to interpret properly. Perhaps the main drawback of this application is phenomenological, as recently emphasized in \cite{Maldacena:2024uhs}. Although we will not address this issue in this paper, it certainly motivates a better understanding of the role the path integral plays in cosmological spacetimes.

\smallskip

The goal of this article is to analyze the application of the gravitational path integral to evaluate the wavefunction of the universe, conditional on future expanding spatial slices having the topology $S^1 \times S^2$. This problem was analyzed first at the classical level by Laflamme \cite{Laflamme:1986bc} and more recently \cite{Anninos:2012ft,Conti:2014uda,Banerjee:2013mca, Maldacena:2019cbz}. This is an interesting example for the following reason. To make the problem tractable, consider a future constant curvature metric on $S^1 \times S^2$. The only free parameter is $L=2\pi R_{S^1}/ R_{S^2}$, which parameterizes the relative proper radius of the circle compared to the sphere (and both are large at late times). In the limit that $L$ is large, the classical analysis predicts a wavefunction
\beq\label{eq:NORMINTRO}
|\Psi(S^1 \times S^2)|^2 \sim \exp{\left( \frac{2\pi \L^2}{3 G_N} \right)},~~~~\text{as }L\to\infty,
\eeq
where $\L$ is the de Sitter radius. We review this calculation in Section \ref{sec:REVIEW}. The expression is independent of $L$, which implies that universes with spatial slices $S^1\times S^2$ give an infinite contribution to probability distributions on future geometries. This is simply because we should integrate over all possible values of $L$, and they all contribute equally for large enough $L$. 

\smallskip

This result is surprising for a few reasons. First of all, having a non-normalizable wavefunction is problematic. Second, it also means that the contribution from $S^1 \times S^2$ spatial universes can become larger than that of $S^3$ universes which, according to Hartle and Hawking, is 
\beq
|\Psi(S^3)|^2 \sim  \exp{\left( \frac{\pi \L^2}{G_N} \right)}.
\eeq
Moreover, there are good reasons to believe that the norm of the wavefunction with $S^1 \times S^2$ spatial slices should be evaluated by the gravitational path integral on $S^2 \times S^2$, where the proper radius of both spheres is $R_{S^2} = \L/\sqrt{3}$. This is the Euclidean section of the Nariai geometry. Similarly, the norm of the Hartle-Hawking wavefunction is expected to be captured by a path integral on $S^4$, although some subtle questions remain regarding this point.\footnote{There is an interesting issue regarding the interpretation of the phase of the partition function on $S^4$  \cite{Polchinski:1988ua,Maldacena:2024spf}. We leave the $S^2 \times S^2$ analysis for a separate publication \cite{WOP_ST}. Related issues are also being considered in \cite{WOP_IM}.}

\smallskip

The problem outlined in the previous paragraph is very similar to the thermodynamic issues that arise in the near-extremal black hole context, mentioned earlier. While the $S^1 \times S^2$ wavefunction becomes independent of $L$ at large $L$, the black hole classical action becomes temperature independent near extremality, and equal to the large classical extremal entropy. In the black hole scenario, this does not lead to divergences (since we never integrate over $\beta$) but does violate a version of the third law of thermodynamics. The resolution is that a specific metric mode that lives near the horizon, which can be described as the length of the throat, becomes light and has to be quantized, such that the classical intuition fails \cite{Ghosh:2019rcj,Iliesiu:2020qvm,Heydeman:2020hhw}. We will refer to those modes as the "Schwarzian modes," since their dynamics is captured by the Schwarzian theory, see \cite{Mertens:2022irh} and references therein. Similarly, in the $S^1\times S^2$ universe, the issue described earlier in this section arises because the classical approximation to the wavefunction becomes independent of the duration of the dS$_2 \times S^2$ inflation period, which emerges in the large $L$ limit. Accordingly, there should be a mode in the cosmological spacetime that requires quantization. 

\smallskip 

This puzzle with the Hartle-Hawking wavefunction on $S^1 \times S^2$ and a resolution was proposed in \cite{Maldacena:2019cbz} based on an enhancement of quantum corrections that appear in the large $L$ limit, also involving Schwarzian modes. The resolution was motivated by the exact quantization of two-dimensional dS$_2$ gravity which was carried out in that paper, see also \cite{Cotler:2019nbi}. The goal of this article is to verify the proposal made in \cite{Maldacena:2019cbz} from the point of view of the full higher-dimensional geometry\footnote{Quantum corrections in the near-Nariai limit are also being considered in \cite{Blacker:2025zca, Maulik:2025phe}. They derive the quantum corrections analytically in the higher-dimensional geometry but restricted to the $dS_2 \times S^2$ throat. In this paper, instead, we numerically consider the full geometry at finite $L$.}. In the black hole case, the analogous goal was achieved in \cite{Kolanowski:2024zrq}. We apply a similar technique to analyze the leading-order quantum corrections that appear in the large $L$ limit for the gravitational path integral evaluating the $S^1 \times S^2$ wavefunction. More specifically, we study off-shell physical metric fluctuations and numerically diagonalize the graviton kinetic operator. We find the presence of modes with an action that vanishes as $1/L$, in accordance with the predictions of \cite{Maldacena:2019cbz}.

\smallskip 

The derivation that we present is valid for finite $L$ and does not rely on any approximation of spacetime in the far past, such as dS$_2\times S^2$, which is valid in a large $L$ expansion. Compared to the black hole scenario, the $S^1 \times S^2$ cosmology case is also complicated by the fact that the geometry is complex, making the analysis more subtle. We discuss the implementation of an appropriate contour for the integration of small fluctuations around a classical solution when the background geometry is complex. Finally, we put all the results together and reproduce the proposal of \cite{Maldacena:2019cbz}, showing that after the inclusion of these quantum corrections the divergence originating from the $L$ integral is regulated. We find the probability distribution is peaked at 
\beq
L \sim \sqrt{ \frac{\pi \L^2}{G_N}}.
\eeq
Including this correction, the norm of the wavefunction is to leading order given by the expression on the right hand side of \eqref{eq:NORMINTRO}. The exponent matches the on-shell action on the Nariai spacetime $S^2 \times S^2$, as expected at least at the classical level, and we leave a discussion on the phase for future work.

\smallskip

Finally, we also consider the analysis of the Hartle-Hawking wavefunction preparing a state in $S^1 \times S^2$, for different choices of spin structures. We apply the same methods as those that were recently applied to supersymmetric black holes \cite{Cabo-Bizet:2018ehj,Iliesiu:2021are} and beyond \cite{Chen:2023mbc}. We find a classical geometry satisfying the Hartle-Hawking no-boundary prescription which asymptotes in the future to a $S^1 \times S^2$ space with periodic fermions along the circle. The classical action as a function of $L$ depends on the spin structure, and for fixed $L$ a universe with periodic fermions is exponentially less likely than one with antiperiodic fermions.  After the inclusion of quantum effects, we find that, nevertheless, the most likely value of $L$ is adjusted such that the norm of the resulting wavefunction is the same regardless of the choice of spin structure.

\smallskip

The organization of the paper is as follows. In \textbf{Section \ref{sec:REVIEW}}, we review the classical analysis of the Hartle-Hawking state, including the standard case of $S^3$ and $S^1 \times S^2$. We emphasize that the classical analysis leads to a divergent norm for the Hartle-Hawking wavefunction of an $S^1 \times S^2$ universe. In \textbf{Section \ref{sec:QuantumCorrections}}, we incorporate quantum corrections. We begin with a general discussion on how the gravitational path integral should be performed on complex spacetimes in \textbf{Section \ref{sec:Formalism}}, and later apply it to the $S^1 \times S^2$ universe in the limit of a large circle in \textbf{Sections \ref{sec:SchwarzianSector}} and \textbf{\ref{sec:rotaionalmodes}}. We reproduce the result suggested in \cite{Maldacena:2019cbz}, working in the full 4d asymptotically de Sitter geometry, and show that quantum effects render the wavefunction normalizable, with a norm that matches the classical approximation to the gravity path integral on $S^2 \times S^2$. In \textbf{Section \ref{sec:Homotopy}}, we analyze the behavior of our results under a complex deformation of the classical background geometry. We find that geometries related by homotopy share the same spectrum of quantum fluctuations and verify it numerically. A summary of the result is in \textbf{Section \ref{sec:assembling}}. In \textbf{Section \ref{sec:othertopo}}, we repeat this analysis for universes with other spatial topologies such as $S^1 \times H^2$ or $S^1 \times S^1 \times S^1$. While the large $S^1$ limit is not relevant in those cases, they serve as useful geometries to further test some of the methods in Section \ref{sec:QuantumCorrections}. In \textbf{Section \ref{sec:wavefunctionspin}}, we consider the Hartle-Hawking wavefunction preparing a state in $S^1 \times S^2$ for different choices of spin structures. We finish with a discussion of open problems and future directions, and leave some technical results for appendices.

\section{The wavefunction of an $S^1 \times S^2$ universe: Classical analysis} \label{sec:REVIEW}

The goal of this paper is to incorporate quantum corrections in the evaluation of the wavefunction of the universe when the spatial topology is $S^1 \times S^2$ instead of the more standard $S^3$ case considered by Hartle and Hawking \cite{Hartle:1983ai}. In order to do so, we begin with a discussion of the classical aspects of this problem. The material in this section is based on the original analysis of Laflamme \cite{Laflamme:1986bc} as well as \cite{Anninos:2012ft,Conti:2014uda,Banerjee:2013mca, Maldacena:2019cbz}. The goal is to review these results and to set up the notation for later.

\subsection{The Hartle-Hawking wavefunction} 
We focus on four-dimensional Einstein gravity with a positive cosmological constant $\Lambda= 3/\L^2$. We can rescale the metric by a factor of $\L^2$ so that the action becomes
\beq
S = \frac{\L^2}{16 \pi G_N} \int \d^4 x \, \sqrt{-g} \, (R-6)  - \frac{\L^2}{8 \pi G_N} \oint \d^3 x \, \sqrt{\gamma} \, K  + \ldots,
\eeq
where $g_{\mu\nu}$ denotes the 4d metric and $\gamma_{ij}$ the metric at the boundary. When discussing proper distances or times, we should recall that the physical metric is $g^{\text{phys.}}_{\mu\nu} = \ell^2 g_{\mu\nu}$. These are convenient conventions to show that our results are automatically valid for any $\ell$. Alternatively, the reader can simply imagine setting $\ell=1$ and then reintroducing it by dimensional analysis. The dots represent matter fields. The results in this section as well as the next one are valid regardless of the details in the matter sector.  

\smallskip 

Let us begin by briefly outlining the no-boundary wavefunction of the universe of Hartle and Hawking \cite{Hartle:1983ai}. The wavefunction of the universe $\Psi[\gamma]$ depends on the choice of spatial metric which we denote by $\gamma_{ij}$ and according to Hartle and Hawking it is given by the gravitational path integral
\beq
\Psi[\gamma] = \int \text{D}g \, \text{D} \Phi\, e^{\i S[g,\Phi] },
\eeq
where the sum is over smooth geometries $g$ that have a single boundary with metric $\gamma$. $\Phi$ is a collective label that represents the path integral over matter fields. Even though this integral is not well defined without a complete theory of quantum gravity, analyzing it in a semiclassical approximation has led to fundamental insights into quantum aspects of black holes, especially in recent years \cite{Lewkowycz:2013nqa, Saad:2018bqo, Saad:2019lba,Almheiri:2019qdq, Penington:2019kki}. The goal of this paper is then to explore the predictions in this regime with
\beq
\Psi[\gamma] \sim \sum_{\text{geometries}} \Psi_{\text{quantum}} \, e^{\i S_{\text{on-shell}}},
\eeq
where the sum is over classical solutions of the equations of motion. The prefactor $\Psi_{\text{quantum}}$ denotes the leading-order quantum corrections to the gravitational path integral around the classical saddle. 

\smallskip 

More concretely, Hartle and Hawking were interested in the wavefunction conditional on the future spatial topology being $S^3$ and, in particular, considered a round constant curvature sphere. The boundary condition relevant to this problem is
\beq
\d s^2 \sim - \d t^2 + \frac{e^{2t}}{4} \, \d \Omega_3^2, 
\eeq
for late times $t$, where $\d \Omega_3^2$ is the round metric on $S^3$. This is simply the four-dimensional de Sitter metric written at late times, in global coordinates. Evaluating the wavefunction at late times then implies that we take $\gamma$ to be the metric of a round sphere with a very large radius proportional to $R_3 \sim \L \, e^{t}/2$. In addition to that, throughout this paper, we also focus exclusively on the expanding branch of the wavefunction, denoted by $\Psi_+$ in \cite{Maldacena:2019cbz}. Since we will only consider this branch in this paper, we will omit the subscript $+$.
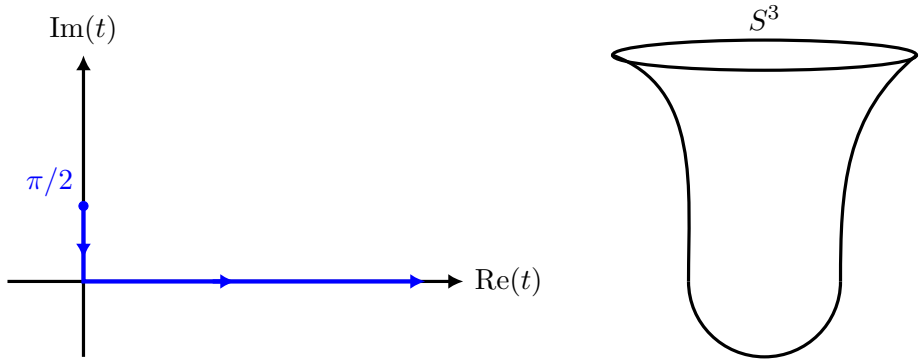
\begin{figure}
    \centering
\begin{tikzpicture}[scale=1]
    \draw[->, very thick,-{Latex[length=2mm, width=2mm]}] (-1,0) -- (5,0) node[right] {$\text{Re}(t)$}; 
    \draw[->, very thick,-{Latex[length=2mm, width=2mm]}] (0,-1) -- (0,3) node[above] {$\text{Im}(t)$};
    \draw[ultra thick, blue,-{Latex[length=2mm, width=2mm]}] (0,1)  -- (0,0.3) ;
    \draw[ultra thick, blue,-{Latex[length=2mm, width=2mm]}] (0,0.5) -- (0,0) -- (2,0);
    \draw[ultra thick, blue,-{Latex[length=2mm, width=2mm]}] (1.7,0) -- (4.5,0);
    \draw[fill, blue] (0,1) circle [radius =0.06] node[above left] {$\pi/2$};
\end{tikzpicture}
~~~~
\begin{tikzpicture}
    \draw[very thick] (1,0) arc (0:-180:1);
    \draw[very thick] (1,0) to[out=90,in=220] (2,3);
        \draw[very thick] (-1,0) to[out=90,in=-20] (-2,3);
        \draw[very thick] (0,3) ellipse (2 and 0.2);
        \draw (0,3.2) node[above] {$S^3$};
\end{tikzpicture}
    \caption{Left: Showing the contour in the complex $t$ plane that corresponds to the Hartle-Hawking no-boundary geometry evaluating the wavefunction on a $S^3$ universe. Right: Picture of the geometry, the Euclidean half-sphere glued to the Lorentzian expanding universe. }
    \label{fig:HHS3}
\end{figure}
\smallskip 

In the semiclassical limit, the relevant geometry discussed in \cite{Hartle:1983ai} consists of the global patch of four-dimensional dS glued at a time reflection symmetry surface to the Euclidean half-sphere that smoothly caps off at the south pole. The geometry can be written as 
\beq
\d s^2 = - \d t^2 +  \cosh^2(t)\, \d \Omega_3^2,
\eeq
where the $t$ contour runs along the real axis and at $t=0$ rotates into $t= - \i \tau$, running from $\tau=0$ until 
 reaching the south pole at $\tau =  \pi/2$. The geometry therefore has no boundary other than the future expanding $S^3$. This is illustrated in Fig.~\ref{fig:HHS3}. A Lorentzian saddle would have, in our conventions, an imaginary action $\i S_{\rm on-shell}$, and would not contribute to the probability distribution $| \Psi[\gamma]|^2$. Due to the incursion into Euclidean space, the Hartle-Hawking saddle is inherently complex, and therefore the action $S_{\rm on-shell}$ has a non-zero real part. In terms of the $S^3$ radius in $\gamma$, which we can parameterize as $R= \L \cosh (t_b)$, we get
 \beq
\Psi(S^3) \sim \text{exp}\Big(\frac{S_{\text{dS}}}{2} - \i \frac{S_{\text{dS}}}{2} \sinh^3(t_b)\Big),~~~~~S_{\text{dS}} = \frac{A}{4 G_N} = \frac{\pi \L^2}{G_N},
 \eeq
 $A$ is the area of the cosmological horizon and $S_{\text{dS}}$ is the dS entropy introduced by Gibbons and Hawking \cite{Gibbons:1977mu}. The probability distribution is $|\Psi(S^3)|^2 \sim \text{exp}(S_{\text{dS}})$.  The phase term comes from the imaginary part of the on-shell action and can depend on the size of the sphere. The leading quantum effects come from the one-loop determinants around this saddle. This leads to small corrections that are logarithmic in the de Sitter entropy, see \cite{Volkov:2000ih} and more recently \cite{Law:2020cpj} and references therein. We will see here that quantum gravity corrections can have more dramatic effects when we generalize this construction.

\subsection{The case with $S^1 \times S^2$ spatial topology} 
\label{sec:S1S2topology}

We shall discuss a modification of the previous analysis by allowing the topology of the future spatial slice to be different. We focus mainly on an expanding universe with spatial topology $S^1 \times S^2$. For simplicity, we take the sphere to be round, and the metric to take a simple product form. Specifically we consider a universe with a late time metric given by
\beq\label{eq:bdycond}
\d s^2 \sim - \d t^2 + e^{2t} \, ( \d x^2 +  \underbrace{\d \theta^2 + \sin^2 \theta \, \d \phi^2}_{\d \Omega^2} ),
\eeq
where the spatial coordinate $x$ has period
\beq
x\sim x + L,
\eeq
and $L$ is a free parameter. The length of the circle at late times is given by $R_{S^1} \sim \L L \cdot e^{t}$ while the radius of the sphere is $R_{S^2} \sim \L \cdot e^{t}$. Therefore, while we are interested in very large $S^1$ and $S^2$ radii, the dimensionless parameter $L$ parametrizes the relative size between the radius of $S^1$ and $S^2$. We will be mainly interested in the dependence of the wavefunction on this parameter $L$. 

\smallskip 

\begin{figure}[t!]
\centering
\includegraphics[width=0.45\textwidth]{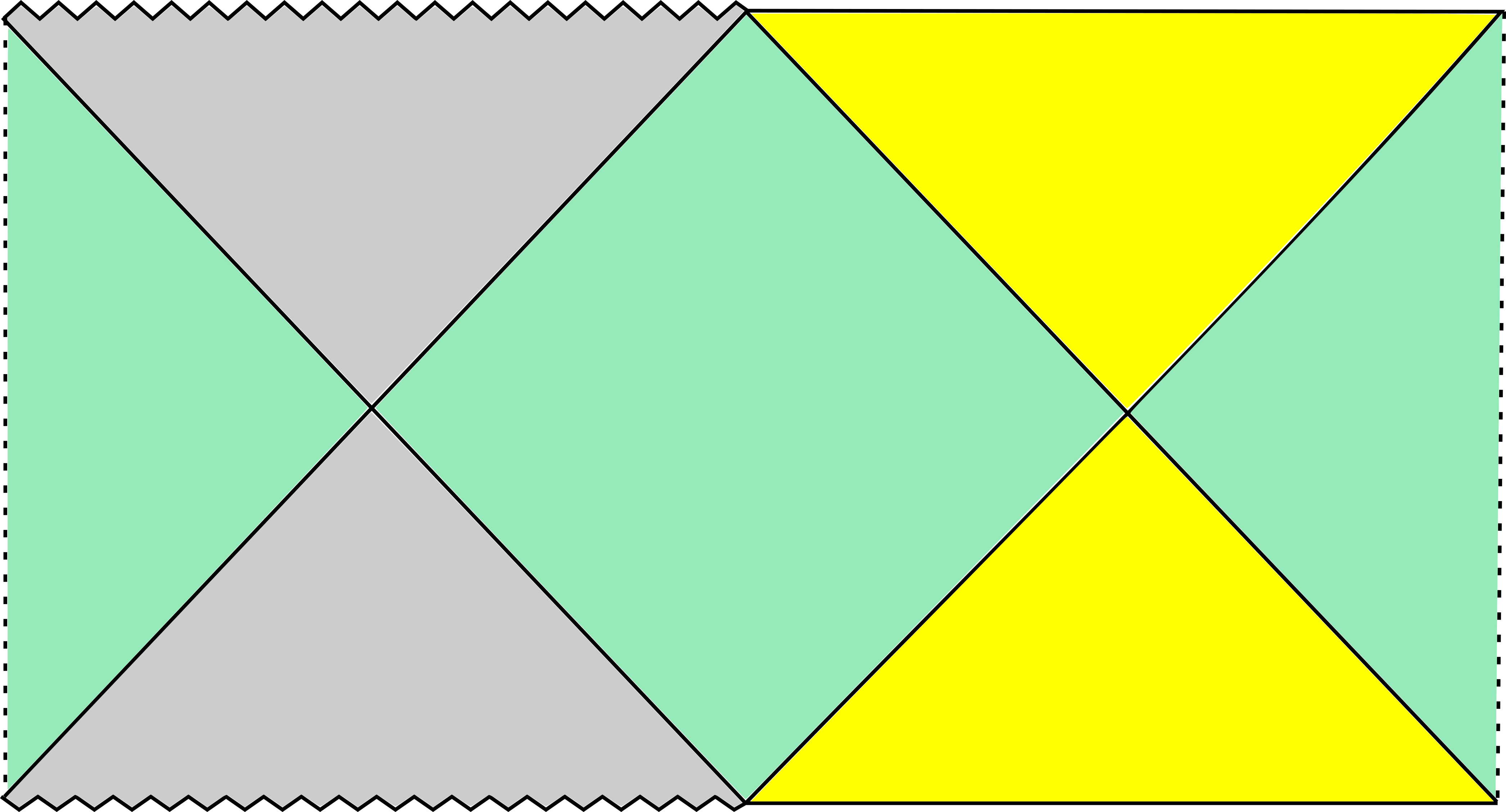}
    \caption{Penrose diagram of a Schwarzschild-dS spacetime. The regions shaded in green correspond to the static patch bounded by the black hole and cosmological horizons, where the borders with dashed lines could be identified or continue endlessly. The grey region is the interior of the black hole, while the yellow region is the expanding patch that we will consider the wavefunction ending on a future $S^1 \times S^2$ spatial slice.}
    \label{fig:Penrose_Diagram}
\end{figure}

Having decided on the boundary conditions in the future, the next step is to determine the possible saddles. In principle, there can be two types of such saddles, one is an orbifold of four-dimensional dS and the second is an analytic continuation of the Schwarzschild-dS black hole \cite{Laflamme:1986bc}. In this paper, we focus on the latter, where quantum gravity corrections are important. The metric is given by the following expression
\beq \label{eq:SdSmetric}
\d s^2 =- \frac{\d \rho^2}{f} + f \d x^2 + \rho^2 \,\d \Omega^2 ,~~~~~~f = \rho^2 + \frac{\mu}{\rho} -1.
\eeq
This metric is precisely the same as that of a Schwarszchild-dS black hole. In that case, the coordinate $x$ would be interpreted as time, while the coordinate $\rho$ would be interpreted as a spatial radial coordinate. The parameter $\mu$ would be related to the mass of the black hole. Following \cite{Laflamme:1986bc,Anninos:2012ft,Conti:2014uda,Banerjee:2013mca, Maldacena:2019cbz}, at large enough $\rho$ in the expanding region, such that $f>0$, the coordinate $x$ becomes a spatial $S^1$ and $\rho$ is the temporal direction, see Fig.~\ref{fig:Penrose_Diagram}. Indeed, at late times we take $\rho\gg \L$ and the metric is
\beq
\d s^2 \sim -  \frac{\d \rho^2}{\rho^2} + \underbrace{\rho^2( \d x^2 + \, \d \Omega_2^2 )}_{=\gamma_{ij} \d x^i \d x^j}.
\eeq
We see that at late times $\rho$ acts as a temporal coordinate. This can be put in the form \eqref{eq:bdycond} if we rewrite it in terms of
\beq 
t = \int \frac{\d \rho }{\sqrt{f(\rho)}}\sim  \log \rho,
\eeq
at late times. 

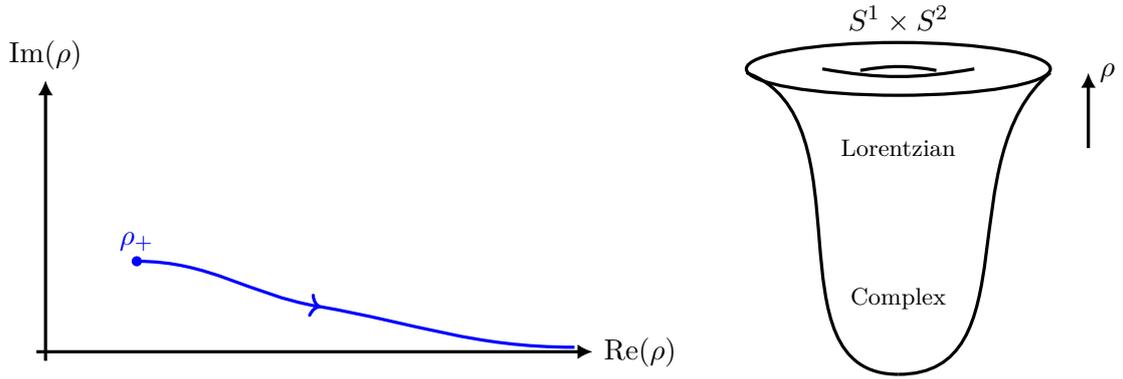
\begin{figure}
\begin{center}
\begin{tikzpicture}[scale=1.2]
    \draw[->, very thick,-{Latex[length=2mm, width=2mm]}] (-0.1,0) -- (6,0) node[right] {$\text{Re}(\rho)$}; 
    \draw[->, very thick,-{Latex[length=2mm, width=2mm]}] (0,-0.1) -- (0,3) node[above] {$\text{Im}(\rho)$};
    \draw[very thick, blue,->] (1,1) to[out=0, in=170] (3.02,0.495);
    \draw[very thick, blue] (3,0.5) to[out=-10,in=180] (5.8,0.05);
    \draw[fill, blue] (1,1) circle [radius =0.05] node[above] {$\rho_+$};
\end{tikzpicture}
~~~~
\begin{tikzpicture}
    \draw[very thick] (0,-1) to[out=0,in=220] (2,3);
        \draw[very thick] (0,-1) to[out=180,in=-20] (-2,3);
        \draw[very thick] (0,3.05) ellipse (2 and 0.35);
        \draw[very thick] (-1,3.05) to [bend right=10] (1,3.05);
        \draw[very thick] (-.5,3.03) to [bend left=10] (.5,3.03);
        \draw (0,3.4) node[above] {$S^1 \times S^2$};
        \draw[-{Latex[length=2mm, width=2mm]},very thick] (2.5,2) -- (2.5,3) node[right] {$\rho$};
        \draw (0,0) node {\footnotesize Complex};
         \draw (0,2) node {\footnotesize Lorentzian};
\end{tikzpicture}
\end{center}
    \caption{Left: Contour in the complex $\rho$ plane corresponding to the no-boundary geometry evaluating the wavefunction on $S^1\times S^2$ universe. The contour starts at $\rho_+$ where the $S^1$ smoothly contracts and ends up along the real $\rho$ axis which is a Lorentzian direction for large enough $\rho$. Right: In the future $\rho$ is time-like and the geometry is real. In the past, the geometry is complex.}
    \label{fig:enter-label}
\end{figure}

\smallskip 

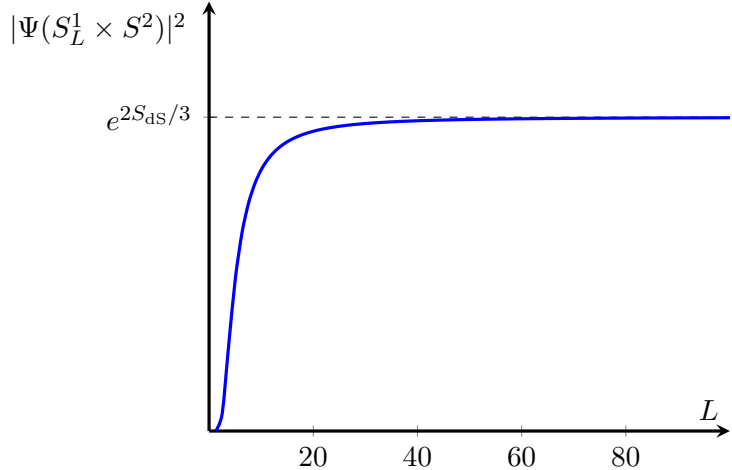
\begin{figure}
    \centering
    \begin{tikzpicture}
          \begin{axis}[axis lines=center, very thick,domain=0:100, y domain 
= 0:9, ymax=10, xtick = {0, 20,40,60,80},
    xticklabels = {0, 20,40,60,80},
    ytick = {0,0.9* 8.12057},
    yticklabels = {$0$, $e^{2 S_{\text{dS}}/3}$}, xlabel={$L$}, ylabel= {\hspace{-3cm}$|\Psi(S_L^1 \times S^2)|^2$\vspace{2cm}}]
            \addplot[samples=80, dashed, thin]
      {0.9*8.12057};
      \addplot[samples=80, smooth, blue, very thick]
      {0.9*exp(2.0944-18.3741/x^2)};
  \end{axis}
    \end{tikzpicture}
    \caption{Plot of the probability distribution $|\Psi(S_L^1 \times S^2)|^2$ as a function of $L$.}
    \label{fig:S1S2prob}
\end{figure}

To complete the description of the solution, we need to determine the parameter $\mu$. The analysis is parallel to that of a black hole. The (future) spatial circle $x$ is contractible at times $\rho_+$ such that $f(\rho_+)=0$. If we take $\rho_+$ as given, then
\beq
\mu = \rho_+ (1-\rho_+^2).\label{eq:murhop}
\eeq
Demanding the solution to be smooth in $\rho_+$ establishes a relation between $\rho_+$, and therefore $\mu$, and the periodicity condition in the future spatial circle $x$. The answer is
\beq \label{eq:periodicity}
L = \pm \i \frac{4\pi}{f'(\rho_+)} = \pm \frac{4\pi \i  \rho_+}{1 - 3 \rho_+^2}.
\eeq
This equation can be solved for $\rho_\pm$ leading to four solutions, namely
\be\label{eq:rhoplusall}
\rho_+=\pm \frac{ 2\pi\i }{3L}  \pm \frac{\sqrt{3 L^2-4 \pi^2}}{3 L}.
\ee
We will see below that the dominant contribution to physical observables arises from the large $L$ regime of this solution (see Fig.~\ref{fig:S1S2prob}). For this reason we will take $L>2\pi /\sqrt{3}$ so that the argument of the square-root is always positive. (The opposite regime of small $L$ was discussed in \cite{Maldacena:2019cbz}.) The four solutions are divided into two pairs, one with a positive imaginary part and another with a negative imaginary part, distinguished by the first sign in \eqref{eq:rhoplusall}. The solutions with $\text{Im}(\rho_+) < 0$ are unphysical; the total proper time elapsed from $\rho_+$ to $\rho \to \infty$ has a positive imaginary part, which implies that there is a negative-time Euclidean evolution. This is problematic since it implies preparation of the matter sector in an unphysical state. Of the two remaining solutions, only one leads to a value of $\mu$, which can be interpreted as mass, with a positive real part. See \cite{Conti:2014uda} and \cite{Maldacena:2019cbz}, and more recently \cite{Hertog:2024nbh}, for further discussion on this point. Following these references, here we assume that the only physical solution is
\be\label{eq:rhoplussaddle}
\rho_+=\frac{ 2\pi\i +\sqrt{3 L^2-4 \pi^2}}{3 L}.
\ee
Combined with \eqref{eq:murhop} this determines the parameter $\mu$ and therefore the complete solution in terms of the parameter $L$. We find that the solution is, in general, complex. (The only exception is the case with $L\to \infty$ which we discuss next.)

\smallskip

Having described the no-boundary geometry, we now present the classical approximation to the wave function. The principle behind the calculation of the on-shell action is quite standard \cite{Hawking:1982dh,Witten:1998zw}, but due to the fact that the saddle is complex, the derivation is more subtle. We carry this out explicitly in the Appendix \ref{app:ACTION}. The wavefunction depends on the relative size between $S^1$ and $S^2$, parametrized by $L$, as well as the total size of any of them, which for concreteness we take to be the circle $L_{proper} = \L \sqrt{f(\rho_b)} L$ and parametrize in terms of time $\rho=\rho_b\to\infty$. The classical approximation to the wavefunction, given by the exponential of the on-shell action, is given by
\beq
\Psi(S_L^1 \times S^2) \sim \text{exp}\Big(-\frac{\pi \L^2}{G_N} \frac{\rho_+^2 (\rho_+^2+1)}{(1-3\rho_+^2)} +\i S_{\text{div}}\Big).
\eeq
where $S_{\text{div}}$ is a term that diverges with the overall proper size of the expanding $S^1 \times S^2$ space, parametrized by $\rho_b\to \infty$, which can be written as an integral on the boundary with local geometric quantities 
\bea \label{eq:Sdiv1}
S_{\text{div}} &=& -\frac{\L^2}{G_N} L (\rho_b^3 - \rho_b/2) \nonumber\\
&=& - \frac{\L^2}{4\pi G_N } \int_{S^1\times S^2} \d^3 x \sqrt{\gamma} \left( 1 - \frac{1}{4} R[\gamma]\right).
\ea
The integral resembles the form of a local counterterm in AdS \cite{Balasubramanian:1999re,Emparan:1999pm}, as explained in \cite{Horowitz:2003he}, but there is no need of additional counterterms here. With $\i S_{\text{div}}$ it shows explicitly that this term leads to an overall phase of the wavefunction. It is useful to rewrite the finite piece solely in terms of $L$ using the specific saddle we are interested in
\be\label{eq:Psifinite}
\Psi_{\text{finite}} \sim \text{exp} \bigg(\frac{\pi \ell^2  }{3 G_N}-\frac{8 \pi^3 \ell^2 }{27 G_N L^2} -\frac{\i \ell^2 \sqrt{3 L^2-4 \pi^2}}{9 G_N}+\frac{ 4 \i \ell^2 \pi^2 \sqrt{3 L^2 -4 \pi^2}}{27 G_N L^2}\bigg).
\ee
Notice that the finite (non-local) piece that depends non-trivially on the relative size $L$ has both a non-vanishing real and imaginary part. When we use it to compute probability distributions, only the real part will remain.

\smallskip

If we evaluate the wavefunction $|\Psi(S_L^1 \times S^2)|^2$, we find that it vanishes at small lengths since $|\Psi(S_L^1 \times S^2)|^2 \sim \exp{\left(- 16 \pi^2 S_{\text{dS}}/(27 L^2)\right)}$ as $L\to 0$ and it reaches a constant at large lengths $|\Psi(S_L^1 \times S^2)|^2 \sim \exp{\left(2S_{\text{dS}}/3\right)}$ as $L\to \infty$. See Fig.~\ref{fig:S1S2prob} for an illustration of how $|\Psi(S_L^1 \times S^2)|^2$ depends on $L$. This result implies that a universe with a large relative circle size $L$, that approaches $\mathbb{R} \times S^2$, is exponentially more likely than a universe with a small $L$. The regime of large $L$ is the Nariai limit which we describe next. In particular, we will explain that this classical intuition is incorrect. This is essential since otherwise the integral over $L$ involved in computing the final probabilities would diverge.

\subsection{The Nariai limit of the $S^1 \times S^2$ wavefunction} 

A special role in this paper is played by the Nariai limit \cite{Nariai, Ginsparg:1982rs, Bousso:1995cc, Bousso:1996au} of the wavefunction of a $S^1\times S^2$ universe. We refer to \cite{Maldacena:2019cbz} for a further discussion of this limit. Here we review some aspects of this limit that will be important for our work. 

\smallskip

At late times, the sizes of both the circle and the sphere become large and proportional to $e^{t}$. The Nariai regime occurs when the dimensionless ratio between the size of $S^1$ and $S^2$ becomes large. Since at late times both radii expand at the same rate, this ratio is time independent and is given by our parameter $L$. Therefore, the Nariai limit corresponds to the large $L$ limit.

\smallskip

As we send the relative size of the circle $L$ to infinity, the location of the horizon becomes
\beq
\rho_+ \sim \frac{1}{\sqrt{3}} + \frac{2\pi \i}{3L} ,~~~~\rho_N = \frac{1}{\sqrt{3}}.
\eeq
The value of $\rho$ at infinite $L$ is called the Nariai radius $\rho_N$. In Euclidean signature, when $\rho$ is close to $\rho_N$ the metric is approximately given by $S^2 \times S^2$ which is called the Nariai solution \cite{Nariai}. In our case, this solution is interpreted as a period of dS$_2 \times S^2$ inflation followed by an asymptotically (locally) dS$_4$ inflation period with spatial $S^1 \times S^2$. Notice that at finite $L$ the solution is complex at early times since $\rho_+$ develops an imaginary part.

\smallskip

At the Nariai radius, the mass parameter becomes
\beq
\mu_N = \rho_N (1 - \rho_N^2) = \frac{2 }{3 \sqrt{3}},
\eeq
and can be interpreted as the maximal mass of a Schwarszchild-dS black hole with a well-behaved Lorentzian metric.

\smallskip

The other solution with $\text{Im}(\rho_+) > 0$ in \eqref{eq:rhoplusall} becomes, at large $L$, $\rho_+ \sim - 1/\sqrt{3}$ and leads to a negative mass. This was one of the reasons for \cite{Maldacena:2019cbz} to argue that such a saddle should not contribute to the gravitational path integral evaluating the wavefunction. In this paper, unless explicitly stated, we focus on the saddle \eqref{eq:rhoplussaddle}, which is unambiguously determined for $L>2\pi/\sqrt{3}$.

\smallskip

We shall analyze the classical approximation to the wavefunction in the Nariai regime. In the large $L$ expansion, the on-shell action from \eqref{eq:Psifinite} becomes
\beq
\Psi(S_L^1 \times S^2) \sim \text{exp}\Big(- \frac{\i \L^2 L}{3 \sqrt{3} G_N} + \frac{\pi \L^2}{3 G_N} + \i \frac{2\pi^2 \L^2}{3\sqrt{3} G_N L} - \frac{8\pi^3 \L^2}{27 G_N L^2} + \ldots \Big),\label{eq:ONSAS1S2}
\eeq
where $\ldots$ are all pure phases. The probability distribution is then
\bea
| \Psi(S_L^1 \times S^2)|^2 &\sim& \text{exp}\Big( \frac{2\pi \L^2}{3 G_N} - \frac{16\pi^3 \L^2}{27 G_N L^2} \Big) \nonumber\\
&\sim& \text{exp}\Big( \frac{2S_{\text{dS}}}{3}  - \frac{16\pi^2 S_{\text{dS}}}{27L^2}\Big).
\ea
This expression is actually exact. It takes a similar form as the partition function of two-dimensional Jackiw-Teitelboim (JT) gravity with a positive cosmological constant \cite{Maldacena:2019cbz, Cotler:2019nbi, Cotler:2024xzz, Dey:2025osp,Nanda:2023wne}. The first term in \eqref{eq:ONSAS1S2} is the analog of energy (proportional to the renormalized length, which in our case is $L$), the second term is analogous to the topological (constant-dilaton) term, and the third is the Schwarzian action inversely proportional to the renormalized length. The $L$-independent term in \eqref{eq:ONSAS1S2} is equal to $A/4G_N$ with $A$ the area of the Nariai horizon. This is not a coincidence since, as explained in \cite{Maldacena:2019cbz}, in the near-Nariai regime our four-dimensional gravity theory becomes well approximated during the $dS_2$ inflation phase by JT gravity.  

\smallskip

It was pointed out in \cite{Preskill:1991tb} that in the near-extremal limit of black holes, their statistical mechanical description breaks down. In that case, the free energy becomes independent of a certain metric mode, which implies that the mode becomes light and has to be quantized at low temperatures. In our case, there is an analogous issue. The absolute value of the wavefunction at large $L$ (analogous to low temperatures) becomes independent of $L$; therefore, it cannot be normalizable at the classical level. The resolution, similar to the black hole case \cite{Ghosh:2019rcj,Iliesiu:2020qvm,Heydeman:2020hhw,Turiaci:2023wrh}, is that we are missing quantum effects that should render the wavefunction normalizable.  (Indeed, the norm should be given by the gravitational path integral on $S^2 \times S^2$.) We will derive this in the next section by analyzing the one-loop corrections to the classical approximation.

\section{The wavefunction of an $S^1 \times S^2$ universe: Quantum corrections} \label{sec:QuantumCorrections}

The focus of this section will be on evaluating quantum corrections around the no-boundary geometry that contributes to the wavefunction of an $S^1 \times S^2$ universe. 

\subsection{The formalism} 
\label{sec:Formalism}
Let us begin with the basic setup. We have found a classical smooth solution that satisfies the appropriate boundary conditions. We called this four-dimensional geometry $g$. The leading quantum effects arise from small fluctuations around this saddle $g_{\mu\nu} \to g_{\mu\nu} + h_{\mu\nu}$. It is useful to define the related metric fluctuation
\begin{equation}\label{eq:tracerevh}
\widetilde{h}_{\mu\nu} = h_{\mu\nu} - \frac{1}{2} g_{\mu\nu} h \,.
\end{equation}
When we integrate over the metric fluctuations $h_{\mu\nu}$, to leading order the action is quadratic on that variable since $g$ satisfies the equations of motion. That quadratic action has a large set of zero-modes due to diffeomorphisms, which we will remove by applying the Fadeev-Popov gauge-fixing procedure.

\smallskip 

We follow \cite{Christensen:1979iy} and use the harmonic gauge. In this gauge, physical configurations satisfy $\nabla^\mu \widetilde{h}_{\mu\nu}=0$. The Fadeev-Popov procedure includes new terms in the action that break diffeomorphism invariance and play the role of imposing the gauge, together with the ghost field that accounts for the path integral measure. The upshot of this analysis is the following correction to the action
\beq
\i S_{\rm gf} = -\underbrace{\frac{\L^2}{32 \pi\, G_N}\,  
\int \d^4 x \, \sqrt{g} \; \nabla^\mu \widetilde{h}_{\mu\sigma} \,\nabla^\nu \widetilde{h}_{\nu}^{~\sigma}}_{\text{gauge fixing term}} - 
\underbrace{\frac{\L^2}{32\pi \,G_N}\, \int \d^4 x \, \sqrt{g} \, \bar{\eta}_\mu\, (-g^{\mu\nu} \nabla^2 - R^{\mu\nu})\,\eta_\nu}_{\text{ghost action}}.
\eeq
The first term in this expression lifts the metric zero-modes due to diffeomorphisms, while the second term over ghost fields $\eta_\mu$ cancels their contribution to the path integral leading to a gauge-fixed finite final answer. As explained in \cite{Kolanowski:2024zrq}, in the presence of matter this gauge should be improved by matter terms to guarantee that gauge modes are orthogonal to physical modes. Since matter fields are turned off in the background, we do not have to worry about including them. 

\smallskip 

After including the gauge-fixing terms and the ghosts, we can expand the action to quadratic order, leading to
\begin{equation}\label{eq:Lopmeasure}
\i S = -\frac{\L^2}{16\pi G_N} \int \d^4 x \sqrt{g} \, \widetilde{h}^{\mu\nu} (\Delta_L h)_{\mu\nu} + \ldots,
\end{equation}
where the dots denote terms that depend on the matter fields or ghosts but not on the metric fluctuations, as well as higher-order terms that do not contribute to the leading-order one-loop analysis. The kernel that appears in the quadratic action for metric fluctuations is \cite{Christensen:1979iy}
\begin{equation}\label{eq:Lichnerowicz_AdS}
\begin{split}
(\Delta_L h)_{\mu\nu} = -\frac{1}{4}\nabla^2 h_{\mu\nu} + \frac{1}{2} \, R_{\rho(\mu}\,h\indices{_{\nu)}^\rho} -\frac{1}{2}\, R_{\mu\rho\nu\sigma}\, h^{\rho \sigma}  
- \left(R_{\sigma(\nu}-\frac{1}{4}\, g_{\sigma(\nu}\, R\right) h\indices{_{\mu)}^\sigma}  - \frac{3}{2} h_{\mu\nu}\,.
\end{split}
\end{equation}
This is the Lichnerowitz operator and has no zero-modes, since the presence of such a mode would signal a failure of the gauge-fixing function. As explained in \cite{Kolanowski:2024zrq} this can happen for a Ricci-flat background, but since we are working with a cosmological constant that issue does not concern us. 

\smallskip 

To perform the path integral over $h$ we need to determine the path integral measure. We follow the standard choice of implicitly defining it via the relation
\begin{equation}\label{eq:ulnorm}
\int {\rm D}h_{\mu\nu} \, e^{-\frac{1}{2}(h,h)} = 1\,, \qquad 
(h,h) = \int \d^4 x \sqrt{g}\,  \widetilde{h}^{\mu\nu} h_{\mu\nu}\,,
\end{equation}
where $(h,h)$ is our choice of ultralocal inner product. One may wonder whether different choices of inner product can affect the final answer; see \cite{Liu:2023jvm}. We restrict ourselves only to this choice of inner product. 

\smallskip 

The result of the path integral is the following. First, diagonalize the operator $\Delta_L$ into the eigenmodes $h^n_{\mu\nu}$ with the eigenvalue $\lambda_n$. The index $n$ here refers to a collective parameter that could be discrete or continuous, depending on the geometry. Next, expand $h_{\mu\nu}$ in such modes
\beq\label{eq:normal}
h_{\mu\nu} = \sum_n c_n h^n_{\mu\nu},~~~~~(h^n_{\mu\nu},h^n_{\mu\nu})=1,
\eeq
where we chose to normalize the eigenmodes. The path integral measure then reduces to a measure of integration over the variables $c_n$ 
\beq
\int \text{D} h \to \int \prod_{n} \frac{\d c_n}{\sqrt{2\pi}}, 
\eeq
and the one-loop path integral is given by
\beq
\int \text{D}h\, e^{-\frac{\L^2}{16 \pi G_N} \int \widetilde{h}^{\mu\nu}(\Delta_L h)_{\mu\nu}} = \text{exp} \left( -\frac{1}{2}\sum_n \log \Big(\frac{\L^2}{16 \pi G_N}\lambda_n\Big) \right).
\eeq
Naively, this is the end of the calculation. We diagonalized the quadratic kernel and evaluated the full path integral. But this is an oversimplification, as we explain next.

\smallskip

Let us first consider a simple case: the gravitational path integral that computes the norm of the Hartle-Hawking wavefunction. Although the classical solution is real, simply a round 4-sphere, there are negative eigenvalues in the quadratic approximation with $\lambda_n<0$. This is a universal feature of the gravitational path integral identified by Gibbons, Hawking, and Perry \cite{Gibbons:1978ac}. This implies that the integral over such modes with negative eigenvalue is divergent. To render the path integral finite, rotate the integration contour of those negative modes. This procedure can lead to a complex result for the path integral \cite{Polchinski:1988ua}. 

\smallskip

Again, this is not the whole story. So far we assumed that the saddle point we expand around is real. Even if the original integral was over real metrics, it can very well be that the saddle point lies somewhere in the complex plane. This is precisely the situation relevant for the wavefunction of the $S^1\times S^2$ universe; although the future boundary conditions are real, the no-boundary proposal forces us to include complex saddles (manifested in the fact that $\rho_+$ or $\mu$ are complex and that we need to take a complex contour in the $\rho$ plane). Most of the formalism above goes through in this case, other than the fact that now the metric is complex, and the operator $\Delta_L$ is not self-adjoint any longer.\footnote{Notice that even though the background geometry is complex, the action involves $(h,\Delta_L h)$ and not $(\bar{h}, \Delta_L h)$. Therefore the relevant condition for $\Delta_L$ to be diagonalizable, i.e. having a complete set of orthogonal eigenmodes with respect to the inner product $(h_1,h_2)$, is just to be a symmetric operator. Of course, since $\Delta_L$ will be a complex operator for complex saddles, its eigenvalues will not be real.}

\smallskip 

Since the operator $\Delta_L$ is no longer self-adjoint, the eigenvalues are no longer real, let alone positive. The resolution is not too different than the case with negative modes, we need to rotate the contour of integration such that the action becomes real. For eigenvalues
\beq
\lambda_n = |\lambda_n| e^{\i \delta_n},
\eeq
the final answer for the one-loop partition function is
\beq\label{eq:compsaddleold}
\int \text{D}h\, e^{-\frac{\L^2}{16 \pi G_N} \int \widetilde{h}^{\mu\nu}(\Delta_L h)_{\mu\nu}} = \text{exp} \left( -\frac{1}{2}\sum_n \log \frac{\L^2|\lambda_n|}{16 \pi G_N} \right) \, \underbrace{e^{-\frac{\i}{2} \sum_n \delta_n}}_{\text{phase}}.
\eeq
The first term on the right-hand side is a positive real number, and the second term is a pure phase. This phase is not important when computing a wavefunction, although we could apply the same formalism to complex geometries that appear in the black hole path integral, for example \cite{Aharony:2021zkr}.

\smallskip

It is interesting to consider what the ghost contribution looks like when the background is a complex metric. If the metric is real, the eigenvalues $\lambda^{\text{ghosts}}_n$ of the ghost determinant $\Delta_{\text{ghosts}}$ can be positive or negative. (We will not need the precise form of this operator in this discussion.) Nevertheless, in the Fadeev-Popov derivation, the path integral over the ghosts is intended to reproduce a Jacobian $| \text{det}\, \Delta_{\text{ghost}}|$ and therefore one is naturally led to consider the absolute value of the eigenvalues $\prod_n |\lambda^{\text{ghosts}}_n|$. Ghosts therefore do not contribute to the phase of the partition function, as emphasized in \cite{Polchinski:1988ua}. When the background metric is complex and, therefore, $\lambda_n^{\text{ghosts}}$ can become complex, the discussion is more subtle. Depending on how one analytically continues the path integral over pure diffeomorphisms when implementing the Fadeev-Popov procedure, the Jacobian is given by the same determinant but up to an overall sign. The sign is uniquely determined in the real case by imposing that the determinant is positive, but this is no longer true if the eigenvalues are complex. Instead, we propose that one should take\footnote{This implies a specific choice of contour of integration when $\lambda_n^{\text{ghosts}}$ is purely imaginary. In our calculation we will not encounter purely imaginary eigenvalues.} $\prod_n (-1)^{\text{sgn}(\text{Re}(\lambda_n^{\text{ghosts}}))+1} \lambda_n^{\text{ghosts}}$. In this way, the analytic continuation is defined to match the answer when the background solutions are real, but generalize otherwise.

\smallskip

We shall now return to the discussion of \eqref{eq:compsaddleold}. This is a reasonable proposal, but why should we rotate the contour this way? To answer this question, we can look at a toy model of a single complex integral, which is rich enough to illustrate the logic
\beq
\int_{\mathcal{C}} \frac{\d z}{\sqrt{2\pi}}\, e^{-N S(z)},
\eeq
and we are interested in the large $N$ limit of this integral with $S(z)$ a fixed analytic function. The contour $\mathcal{C}$ could be on the real axis. Consider then a function such that the saddle points $S'(z_*)=0$ are in the complex plane, or more generally $z_*  \notin \mathcal{C}$. The steepest descent method indicates that we should rotate the contour into a new one $\mathcal{C}'$ that passes through $z_*$ and such that $\text{Im}[ S(z)]$ is constant. This contour deformation might not pick all the saddle points. This is a difficult question in the gravity context; the only goal here is to decide how a given saddle contributes to the integral, assuming the steepest descent contour passes by it. Locally, the integrand near $z_*$ will behave as 
\beq
S(z) = S(z_*) + \frac{1}{2} S''(z_*) (z-z_*)^2 + \ldots,
\eeq
where the dots denote higher-order terms away from the saddle point. The second derivative need not be real, and we denote the phase by $S''(z_*) = |S''(z_*)| e^{\i \delta(z_*)}$. Locally, the steepest descent contour is
\beq
z = z_* + e^{-\frac{\i}{2} \delta(z_*)} \tilde{z},
\eeq
where the parameter $\tilde{z}$ is real in this approximation. The Gaussian integral along the steepest descent contour is then an integral of the real variable $\tilde{z}$. This makes the $\tilde{z}$ dependent part of the action to be real, but introduces an overall phase from the change of variables from $z$ to $\tilde{z}$. Therefore, the large $N$ approximation to the original integral is
\beq
\int_{\mathcal{C}} \frac{\d z}{\sqrt{2\pi}}\, e^{-N S(z)} \approx \sum_{z_*}  \frac{e^{-\frac{\i}{2} \delta(z_*)}}{\sqrt{|S''(z_*)|}}\, e^{-N S(z_*)},
\eeq
where the sum is over those $z_*$ that are encountered by a steepest descent contour deformable to $\mathcal{C}$. Generalizing this analysis to the gravitational path integral leads directly to our formal answer \eqref{eq:compsaddleold}. In this paper, we apply this method to the complex saddle that contributes to the no-boundary wavefunction preparing $S^1 \times S^2$, and for the reasons explained above we will focus on the absolute value of the eigenvalues. 

\subsection{Spherically-symmetric modes: Schwarzian sector}
\label{sec:SchwarzianSector}

Here we apply the above formalism and evaluate the leading quantum gravity corrections to the geometry that prepares the no-boundary state $\Psi(S_L^1 \times S^2)$. In particular, we are interested in identifying and characterizing the presence of so-called Schwarzian modes that are physical and off-shell, and whose action vanishes in the large $L$ limit. These will arise from the presence of an early period of dS$_2 \times S^2$ inflation, and give the dominant quantum effect, as anticipated in \cite{Maldacena:2019cbz}. In order to achieve this, we numerically diagonalize the Lichnerowicz operator in the full asymptotically dS$_4$ geometry at finite $L$ and identify the lowest-lying modes. The analysis closely follows the recent work \cite{Kolanowski:2024zrq}, although the complex (and, at late times, Lorentzian) nature of the geometry makes the analysis more involved in interesting ways.

\smallskip

We first discuss what we anticipate from the numerical analysis. To analyze the spectrum of the Lichnerowicz operator, we first Fourier expand $h_{\mu \nu}$ in eigenfunctions of the Killing isometry $\partial_x$ of the background in the momentum $$k_n=\frac{2\pi n}{ L}.$$ We will refer to the index $n$ as ``momentum'' as well. In the end, working in a sector of fixed momentum $k_n$, we would get a quadratic generalized eigenvalue problem, and we expect to see a set of nearly zero modes with eigenvalue linear in $\lambda_n \propto k_n $, therefore becoming exact zero modes as $L \to \infty$. Furthermore, as first observed in \cite{Maldacena:2019cbz} by analyzing the Schwarzian couplings in the wavefunction, the eigenvalues should be purely imaginary. We will also study the eigenvectors to demonstrate that they are indeed localized when $L$ is large.

\smallskip

The large $L$ near-Nariai limit corresponds to a natural near-extremal limit of the Schwarzschild-de Sitter geometry, without the need of introducing gauge fields. Hence, the analysis is much simpler compared to the AdS Reissner-Nordstr\"om geometry studied in \cite{Kolanowski:2024zrq}. The issue there is that $\U(1)$ modes do not decouple from the metric modes in the full geometry. However, our setup is complicated by the complex nature of the solution, as we will see below. We also anticipate that the Schwarzian modes are homogeneous in the unit sphere and therefore preserve the background SO(3) rotation symmetry. 

\smallskip

Motivated by these considerations, we consider the following simple ansatz for the eigenfunctions of the Schwarzian modes \cite{Kolanowski:2024zrq}
\be
h_{\mu \nu} \d x^\mu \d x^\nu= e^{\i k_n x} \bigg(f_1(\rho)  f(\rho)\d x^2+f_2(\rho)\frac{\d\rho^2}{f(\rho)}+2\i f_3(\rho)\d x \d\rho+\rho^2 f_4(\rho) \d \Omega^2_2   \bigg),
\ee
where the functions $f_i (\rho)$ with $i \in \{1,\ldots,4 \}$ that only depend on $\rho$ would generally be complex. We will only consider modes that are normalizable according to the ultralocal norm \eqref{eq:normal}. As explained in \cite{Kolanowski:2024zrq}, in the absence of matter, we expect the Schwarzian modes to satisfy the traceless-transverse gauge, where we impose
\be
g^{\mu \nu}h_{\mu \nu}=0, \quad \nabla^\mu h_{\mu \nu}=0. 
\ee
This constraint will fix all gauge freedom for $k_n \neq 0$, and allow us to recast the question into a single second-order ODE with a scalar function we call $u(\rho)$ below. 

\smallskip

The traceless condition allows us to eliminate $f_4$
\be
g^{\mu \nu}h_{\mu \nu}=0 \implies f_4=\frac{1}{2} (f_1-f_2).
\ee
The $\rho$-component of the transverse condition allows us to further eliminate $f_1$ 
\be
\nabla^\mu h_{\mu \rho}=0 \implies f_1=\frac{1}{2f-\rho f'} \bigg[f_2(\rho f'+6f)+2 \rho f f'_2+\frac{4 n \pi \rho f_3}{L} \bigg],
\ee
and we substitute $f_1$ into the $x$-component of the transverse equation $\nabla^\mu h_{\mu x}=0$. This gives an equation involving $f_2, f_2', f_3$, and $f'_3$.
In order to express $f_2$ and $f_3$ in terms of a single scalar function $u(\rho)$, we assume an ansatz
\be
f_2(\rho)=u(\rho)+H(\rho) f_3(\rho),
\ee
then to eliminate any dependence on $f_3'(\rho)$, we choose 
\be
H(\rho)=\frac{ L (2f-\rho f')}{4 n \pi \rho},
\ee
and then
\be
f_3(\rho)=-\frac{4  n \pi  L [u(\rho f'+6f)+2 f \rho u')]}{\rho [16 n^2 \pi^2+ L^2(f'^2-2f f'')]}. 
\ee
Therefore, $f_2$ and $f_3$ can be expressed in terms of a single scalar function $u(\rho)$. Now we are ready to consider the spin-2 Lichnerowicz operator equation
\be
\Delta_L h_{\mu \nu}=\lambda h_{\mu \nu}.
\ee
The operator equation can be systematically reduced to a generalized eigenvalue problem for $u(\rho)$ with the following second-order ODE
\be
A(\rho)u''(\rho)+B(\rho)u'(\rho)+C(\rho)u(\rho)=\lambda G(\rho) u(\rho).
\ee
The explicit form of the ODE is not very illuminating. Instead, we will solve the problem via collocation methods, specifically with the Chebyshev-Gauss-Lobatto grids \cite{doi:10.1137/1.9780898719598, Grandclement:2007sb, Dias:2015nua}\footnote{We would like to thank Maciej Kolanowski for useful discussions about the implementation of this method.}.

\smallskip

At this point, we need to specify the contour in the $\rho$ plane that determines the precise background geometry, going from $\rho_+$ (where we implement the no-boundary prescription) toward the future $\rho \to \infty$. We make the following concrete choice
\be \la{eq:contour}
\rho(z)=\frac{1}{2}\rho_+ (1-z)+|\Re[\rho_+]| \bigg(\frac{1+z}{1-z} \bigg).
\ee
where the domain of the parameter $z$ is $z \in [-1, 1]$. The fact that it is compact will facilitate the numerical analysis of the ODE for $u$ based on the collocation methods. (We include $|\Re[\rho_+]|$ in the second term in order not to spoil the fact that we are moving toward positive real axis as we increase $z$.) In Section \ref{sec:Homotopy}, we will revisit this choice and determine whether and how the quantum corrections depend on the choice of contour.

\smallskip

Performing the map \eqref{eq:contour}, we have an ODE in terms of $u(\rho(z))\to u(z)$, ensuring that the boundaries $z=\pm 1 $ are at least regular singular points. In terms of the ODE, it means  $\frac{B(z)}{A(z)}$ has at most a simple pole, while $\frac{C(z)-\lambda G(z)}{A(z)}$ has at most a pole of second order at these points. Thus, we can employ the Frobenius series method to analyze these endpoints. For the solutions to preserve the near-horizon and asymptotic structures, we perform a local analysis with the Frobenius series solutions. For the asymptotic region $z=1$ we take
\be
u(z) =  (1-z)^s \sum_{p=0}^\infty \mathcal{A}_p (1-z)^p,
\ee
where the leading order term around $z=1$ with $p=0$ is the indicial equation and has to be zero
\be\label{eq:lambdabound}
s^2-7s+10-4 \lambda=0 \implies s_\pm=\frac{1}{2} (7 \pm \sqrt{9+ 16 \lambda}).
\ee
This behavior is consistent with a similar analysis done for the hyperbolic AdS black hole example studied in \cite{Kolanowski:2024zrq}, with a change in the sign of $\lambda$ since we are in dS. This method works as long as $\lambda \notin (-\infty, -9/16]$. This does not happen in our case, since $\lambda$ will naturally be complex.\footnote{An exception would be a spatial topology $S^1 \times H^2$ to be analyzed in Section \ref{sec:othertopo}, where we expect the eigenvalues to be real from analyzing the Schwarzian couplings. Here with $S^1 \times S^2$, we expect $\lambda$ to be imaginary to leading order in the large $L$ limit.} Since the solutions must preserve the asymptotic Dirichlet boundary conditions, as long as we have sufficiently fast decay, we could pick $s=\frac{5}{2}$ such that we only have one independent solution. 

\smallskip
 
Similarly, near the other end of the geometry $z=-1$ we expand
\be
u(z) =  (1+z)^{s'} \sum_{p=0}^\infty \mathcal{A}_p (1+z)^p,
\ee
the indicial equation gives
\be
n^2-4 (1+s')^2=0 \implies s'_\pm=\pm \frac{ n }{2}-1.
\ee
Again, this is consistent with the near-horizon behavior observed in \cite{Kolanowski:2024zrq} for the AdS black holes, as it should be for the near-horizon Rindler structure. A factor of two difference in the indicial parameter comes from the different behaviors in the choices of contour. Near $z=-1$ expansion of \eqref{eq:contour} is
\be
\rho \approx \rho_++\frac{1}{2}( |\Re[\rho_+]|-\rho_+)(1+z)+\ldots,
\ee
while the map considered in \cite{Kolanowski:2024zrq} would have a next-to-leading-order piece $\# (1+z)^2$. The solutions will need to be regular at the horizon, therefore, we will stick to the faster decay with $s'_+$ and impose $|n| \geq 2$. The norm of $n=0, \pm 1$ would diverge and be non-normalizable. This is precisely what we expect from an independent path integral derivation \cite{Maldacena:2019cbz}, which we can now confirm in the full geometry.

\smallskip

We motivate the following substitution 
\be
u(z)=(1-z)^{\frac{5}{2}}(1+z)^{\frac{|n|}{2}-1}w(z),
\ee
along with the physical saddle identified in \eqref{eq:rhoplussaddle} for $\rho_+$ in terms of $L$. In order to track the lowest set of eigenvalues of $\Delta_L$ for a fixed momentum mode $n$, we can write the equation as a generalized eigenvalue problem
\be
\mathcal{D} w(z)= \lambda_n \mathcal{G} w(z),
\ee
where
\be
\mathcal{D}=A(z) \partial^2_z+B(z) \partial_z+C(z), \quad \mathcal{G}=G(z),
\ee
that can be put in Mathematica's $Eigensystem[\{ \mathcal{D}, \mathcal{G} \}]$ to solve the corresponding eigenvalues $\lambda_n$. We discretize the interval $z \in [-1,1] =[z_-, z_+]$ into Chebyshev-Gauss-Lobatto grid with $N+1$ points by \cite{doi:10.1137/1.9780898719598, Grandclement:2007sb, Dias:2015nua}
\be
z_j=\frac{z_+ + z_-}{2}+\frac{z_+-z_-}{2} \cos{\bigg( \frac{\pi j}{N} \bigg)}, \quad j=0, 1,...,N.
\ee
Note that with this construction, the grid points actually start with a reverse order: $z_j=[1,...,-1]$. This is important when imposing boundary conditions at $z=1, -1$. We also construct the differentiation matrices as follows
\be
D^{(1)}_{jj}=\sum_{k \neq j}\frac{1}{z_j-z_k}, \quad D^{(1)}_{ij}=\frac{a_i}{a_j} \frac{1}{z_i-z_j} \quad (i \neq j), \quad a_j= \prod_{k \neq j} (z_j-z_k),
\ee
and $D^{(2)}=(D^{(1)})^2$. This algorithm could reach an exponential accuracy with respect to the number of grid points $N$; however, the computational time complexity increases as $\mathcal{O}(N^3)$. In order to probe very small values of $1/L$, it is generally required to increase the number of grid points and to go beyond the machine precision.

\smallskip

Then we evaluate the second-order ODE on these grid points $\vec{z}$ with
\be
\mathcal{D}=A(\vec{z}) D^{(2)}+ B(\vec{z}) D^{(1)}+C(\vec{z}), \quad \mathcal{G}=G(\vec{z}),
\ee
where we note $A(\vec{z}),  B(\vec{z}), C(\vec{z}), G(\vec{z})$ must first be diagonalized with the diagonal elements given by the corresponding functions evaluated at each grid point value from $\vec{z}$. We impose the Dirichlet boundary condition at $z=1$, where we simply exclude the first row and column of the two matrices ${\mathcal{D}, \mathcal{G}}$. At $z=-1$, we only need to make sure that the matrices are regular. This is indeed the case as we found the function $A$ vanishes at $z=-1$, while $B$, $C$, and $G$ are regular there.

\smallskip

\begin{figure}[t!]
  \centering
  \begin{subfigure}[b]{0.6\textwidth}
    \centering
    \includegraphics[width=\textwidth]{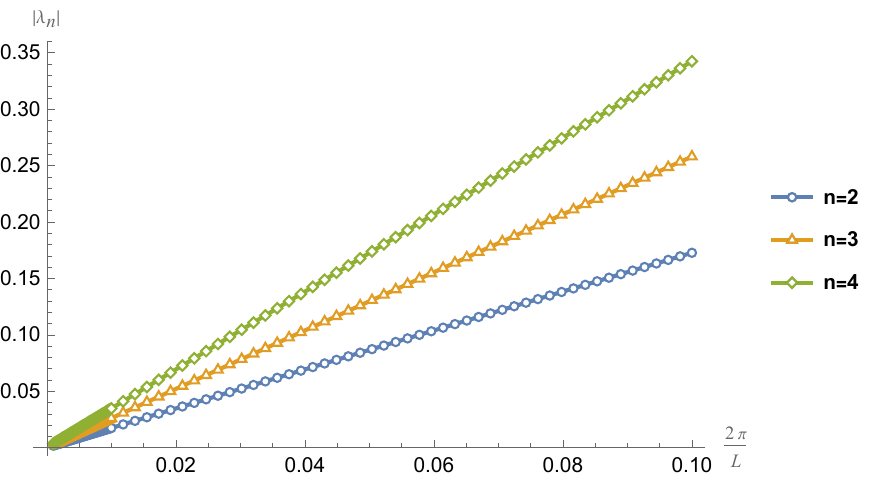}
    \caption{a: The eigenvalue spectrum over an extended range of $1/L$.}
    \label{fig:t10fullfit}
  \end{subfigure}
  
  \vspace{0.5cm}

  \begin{subfigure}[b]{0.45\textwidth}
    \centering
    \includegraphics[width=\textwidth]{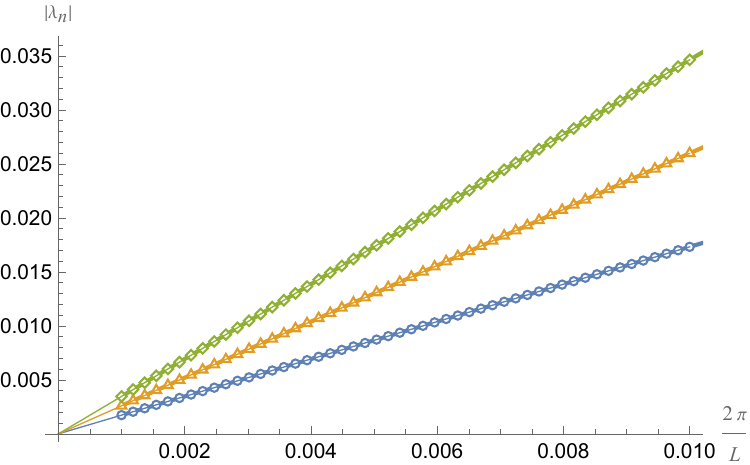}
    \caption{b: The eigenvalue spectrum at small $1/L$.}
    \label{fig:t10fit_zoomin}
  \end{subfigure}
  \hfill
  \begin{subfigure}[b]{0.45\textwidth}
    \centering
    \includegraphics[width=\textwidth]{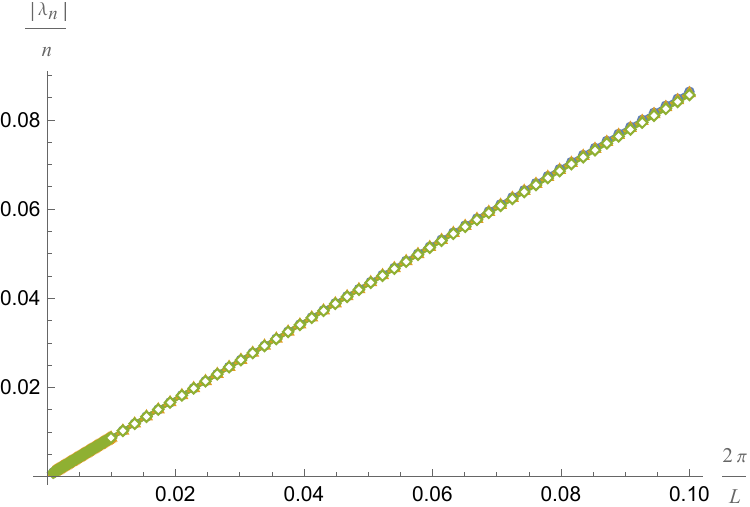}
    \caption{c: Rescaled eigenvalues.}
    \label{fig:t10rescale}
  \end{subfigure}
  
  \caption{The eigenvalue spectrum of the full cosmological spacetime, along with the fit \eqref{eq:quadraticfit} demonstrating the linear behavior at small values of $1/L$. Here we sample $50$ points for $\frac{2 \pi}{L} \in [0.001,0.01]$ and $50$ points for $\frac{2 \pi}{L} \in [0.01,0.1]$. In Fig.~\ref{fig:t10rescale}, we rescale with $\frac{|\lambda_n|}{n}$ and found that the eigenvalues for the three momenta modes are overlapping, showing that $|\lambda_n| \propto \frac{n}{L} $ at large $L$.}
  \label{fig:t10eigvalues}
\end{figure}

The eigenvalues for the Schwarzian modes with the lowest three positive momentum modes $(n=2,3,4)$ as a function of $1/L$ are shown in Fig.~\ref{fig:t10eigvalues}. We consider a quadratic polynomial fit for $k_n/n=2\pi/L \in [0.001, 0.01]$ 
\be \la{eq:quadraticfit}
|\lambda_n|=a_n+b_n \,k_n+c_n \, k_n^2 + \mathcal{O}(k_n^3),
\ee
where we find
\be
a_2 \sim \mathcal{O}(10^{-8}), \quad b_2 \approx 0.8660, \quad c_2 \approx -0.0034.
\ee
\be
a_3\sim \mathcal{O}(10^{-7}), \quad b_3 \approx 0.8657, \quad c_3 \approx -0.0057.
\ee
\be
a_4\sim \mathcal{O}(10^{-7}), \quad b_4 \approx 0.8661, \quad c_4 \approx -0.0079.
\ee
The precise numerical values could depend weakly on the number of grid points and precision we keep. Nevertheless, this is in agreement with the expectation that the modes become zero modes as we approach extremality at large $L$, and they scale linearly with $k_n$ that is proportional to $1/L$. See Fig.~\ref{fig:t10eigvalues} with the fit. Furthermore, we see in Fig.~\ref{fig:t10rescale}, the slope of the eigenvalues scales linearly with $n$, as we can also see from the fact that $k_n\propto n$ and $b_n$ is approximately $n$-independent $b_{2,3,4} \sim 0.866$, with the small differences being an artifact of only using a quadratic fit. The result is expected since they should be identified with the Schwarzian mode lifted onto the full geometry.

\smallskip

\begin{figure}[t!]
  \centering
\begin{subfigure}[b]{0.45\textwidth}
    \centering
    \includegraphics[width=\textwidth]{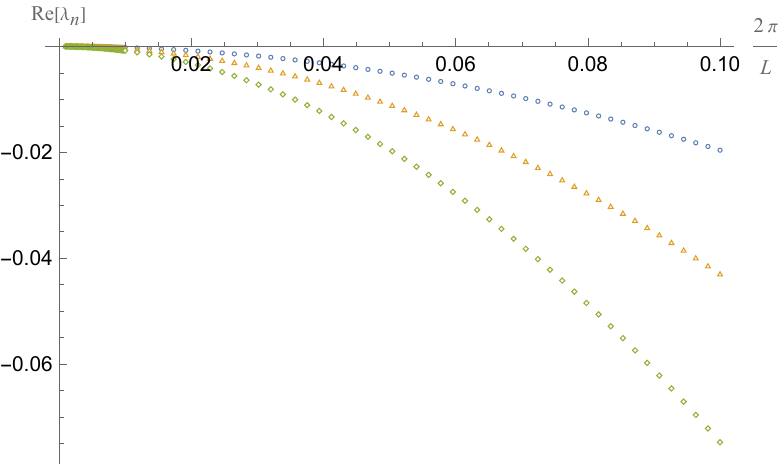}
    \caption{a: The real part of $\lambda_n$.}
    \label{fig:t10Re}
  \end{subfigure}
  \hfill
  \begin{subfigure}[b]{0.45\textwidth}
    \centering
    \includegraphics[width=\textwidth]{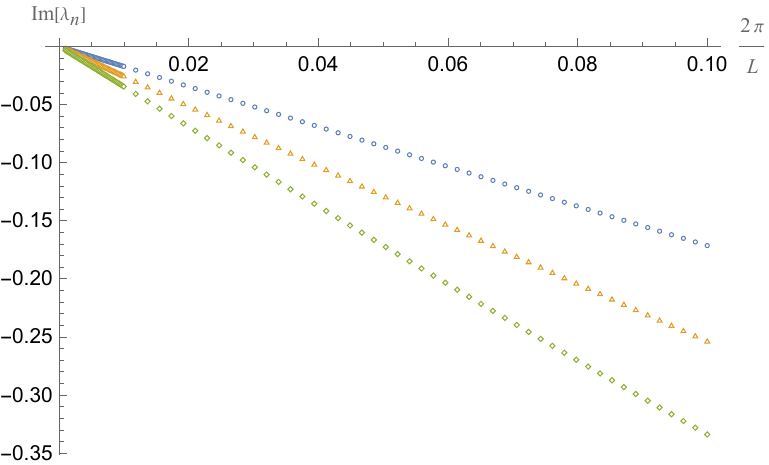}
    \caption{b: The imaginary part of $\lambda_n$.}
    \label{fig:t10Im}
  \end{subfigure}

    \vspace{0.5cm}

    \begin{subfigure}[b]{0.6\textwidth}
    \centering
    \includegraphics[width=\textwidth]{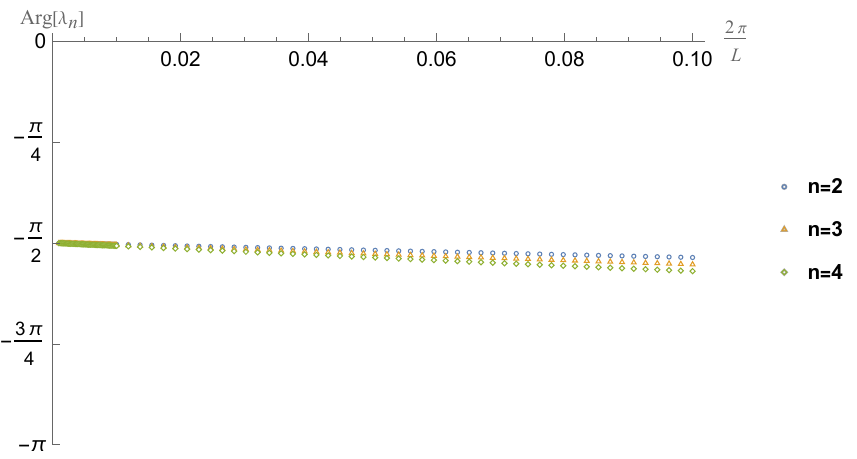}
    \caption{c: The phase of $\lambda_n$.}
    \label{fig:t10phase}
  \end{subfigure}
  
  \caption{Top panel: we plot the real and imaginary parts of the eigenvalues $\lambda_n$ for the three momenta modes over the extended range of $1/L$. Bottom: we plot the phase of $\lambda_n$, indicating that the eigenvalues are dominated by the imaginary part at large $L$.}
  \label{fig:t10eigen}
\end{figure}

Note that here we are taking the absolute value of $\lambda_n$ since they are complex, as shown in Fig.~\ref{fig:t10eigen}. As discussed in Section \ref{sec:Formalism}, we only need $|\lambda_n|$ and will not discuss the overall phase in the one-loop determinant. However, from Fig~\ref{fig:t10phase}, we see that the $\text{Arg}[\lambda_n] \approx -\frac{\pi}{2}$ across the whole parameter regime of $1/L$ for the three momenta modes we studied. This indicates that at large $L$, the eigenvalues are imaginary, consistent with the fact that the Schwarzian coupling for dS$_2 \times S^2$ is imaginary \cite{Maldacena:2019cbz}. Given our convention for the modes $e^{\i k_n x}$, the sign of the imaginary part is also consistent with the prediction from the Schwarzian theory.

\smallskip

\begin{figure}[t!]
\centering
\includegraphics[width=0.45\textwidth]{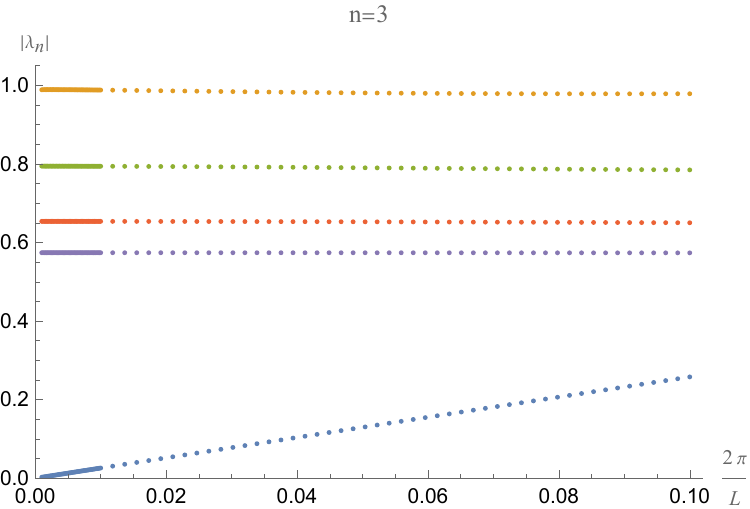}
\caption{We plot the 5 lowest eigenvalues $|\lambda_n|$ for $n=3$ as an example. The lowest one, corresponding to the Schwarzian mode, scales linearly with $1/L$ at large $L$, as we have demonstrated already in Fig.~\ref{fig:t10eigvalues}. However, the higher eigenvalues are independent of both $L$ and $n$. Note that if we continue the plot for $\frac{2 \pi}{L}>0.1$, the eigenvalue branches can cross.}
\label{fig:HigherEig}
\end{figure}

In Fig.~\ref{fig:HigherEig}, we could also see the distinction between the eigenvalue branch corresponding to the Schwarizan modes with other eigenvalue branches. In fact, these other branches are independent of both $L$ and $n$ in the large $L$ limit, as expected.

\smallskip

\begin{figure}[t!]
\centering
\includegraphics[width=0.4\textwidth]{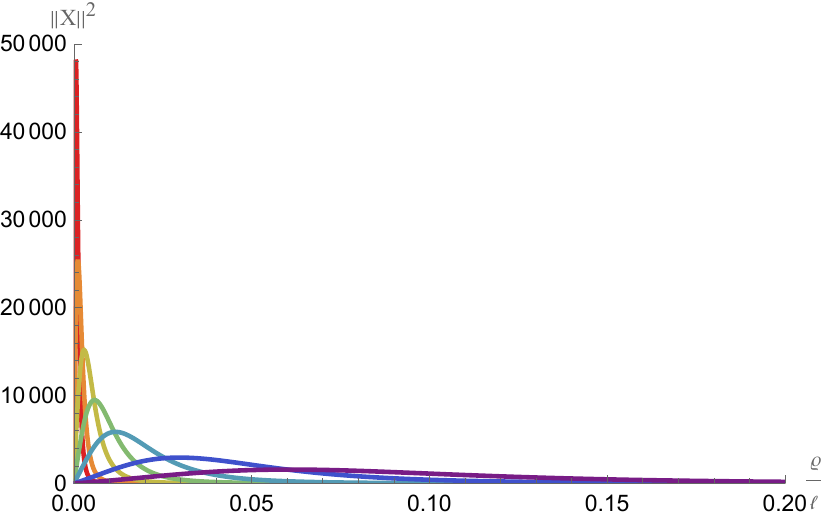}
\includegraphics[width=0.4\textwidth]{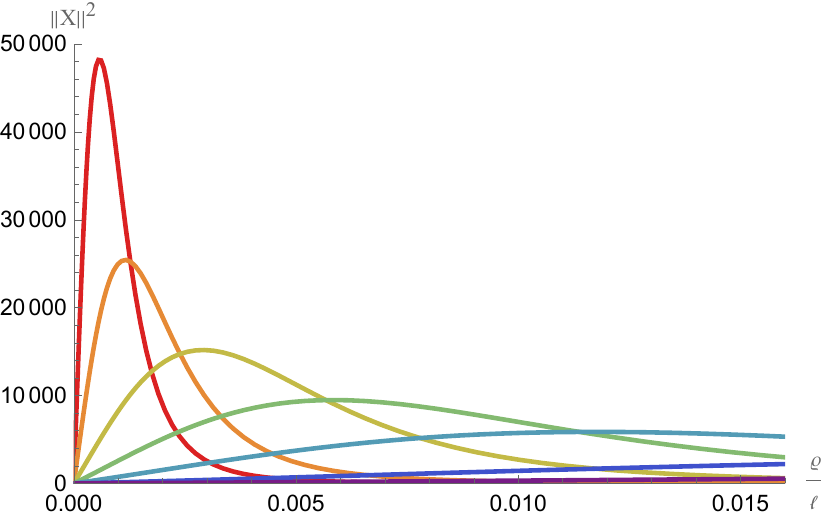}
\caption{The norm defined as the absolute value of the integrand in \eqref{eq:normSch} over the proper time $\varrho=\ell|\ln{(\rho/\rho_+)}|$ for $n=3$ with $\frac{2 \pi }{L} \in [0.1, 0.05,  0.02, 0.01, 0.005, 0.002, 0.001]$. We see that as we lower $1/L$, the norm becomes increasingly localized toward the horizon, and the peak amplitude increases. Note here we have to work with a larger number of grid points compared to the evaluation of the eigenvalues.}
\label{fig:normSch}
\end{figure}

We can also study the eigenvectors and consider the norm of the perturbation
\bea \la{eq:normSch}
||h||^2&=&\int \d^4 x \sqrt{-g} \tilde{h}^{ \mu \nu}h_{\mu \nu}
\no\\
&=&\int_{\rho_+}^\infty \d \rho \bigg[ 4 \pi L \rho^2 (f_1^2+f_2^2 -2 f_3^2+2 f_4^2)\bigg].
\eea
We plot the absolute value of the integrand given above over the proper time, this is shown in Fig.~\ref{fig:normSch}. The fact that the norm is localized near the horizon at large $L$ is what we expect from the Schwarzian modes.

\FloatBarrier

\FloatBarrier

\smallskip

In summary, we found graviton modes that, as $L$ increases, are localized in the far past during the dS$_2 \times S^2$ inflation period. They also have an action that, when properly normalized with respect to the ultralocal measure, is proportional to their momenta $k_n$ as well as $\L^2/G_N \sim S_{\text{dS}}$. Their contribution to the quantum correction of the wavefunction is
\beq
\Psi_{\text{quantum}} \sim \frac{1}{\prod_{n\neq\{0,\pm1\}} ( S_{\text{dS}} k_n)^{1/2} } \sim \frac{1}{S_{\text{dS}}^{3/2} L^{3/2}},
\eeq
where we have not kept track of the overall $L$-independent prefactor (including a phase) arising from all other modes that have a finite action at $L=\infty$. In particular this can introduce other powers of $S_{\text{dS}}$ as well. We see that the quantum corrections are increasingly relevant as the relative size of $S^1$ compared to $S^2$ becomes larger. 

\smallskip

Is there an effective action describing the dynamics of these light modes? The answer is provided by the Schwarzian action
\beq
\i S = \Phi_r  \int \d x\,  \{ F(x), x\}, ~~~~\Phi_r = \i \frac{\mu_N S_{\text{dS}}}{2\pi}.
\eeq
The action of the Fourier modes of the field $F(x)$ matches the action of the nearly zero-modes we found above in the full geometry \cite{Maldacena:2019cbz}. The quadratic expansion of this action has all the same properties as the modes we found from the four-dimensional analysis, as long as $\Phi_r$ is imaginary. The precise coefficient $\Phi_r$ can be matched by comparing with the $1/L$ term in the large $L$ expansion of the wavefunction of the universe \eqref{eq:ONSAS1S2}. The full non-linear completion can be derived via a dimensional reduction of the geometry to dS$_2 \times S^2$ in a large $L$ expansion, see \cite{Maldacena:2019cbz}. As stressed earlier, in this paper we have reproduced this from a finite $L$ analysis of the full four-dimensional geometry. 

\smallskip

It would be interesting to understand the fate of the other unphysical saddles identified in \eqref{eq:rhoplusall}. We could denote the four saddles in \eqref{eq:rhoplusall} as $(++), (-+), (+-)$, and $(--)$ based on their signs of imaginary and real parts of $\rho_+$ at large $L$, respectively. Then the physical saddle is $(++)$. In all four saddles we could identify the lowest set of eigenvalues corresponding to the Schwarzian modes with the same properties described in this section.\footnote{A caveat is that the four saddles are distinct as we are always picking a contour from $\rho_+$ to positive real infinity. Had we picked a contour that goes to negative real infinity for $(+-)$ and $(--)$, then they are in fact the same saddles as $(-+)$ and $(++)$, respectively.} In fact, the eigenvalues of the $(-+)$ and $(+-)$ saddles turn out to be exactly the complex conjugate of the physical saddle $(++)$, while the eigenvalues of $(--)$ saddle are the same as $(++)$. This is expected since the $(-+)$ saddle is the conjugate of the physical saddle, while the $(+-)$ is the solution generated from \eqref{eq:periodicity} in the other conjugate branch. It indicates the Schwarzian modes that are localized in the dS$_2$ throat are more universal from the near-horizon behaviors, but nevertheless saddles other than $(++)$ are unphysical and should be excluded from gravitational path integral.

\smallskip

\begin{figure}[hbt!]
\centering
\includegraphics[width=0.45\textwidth]{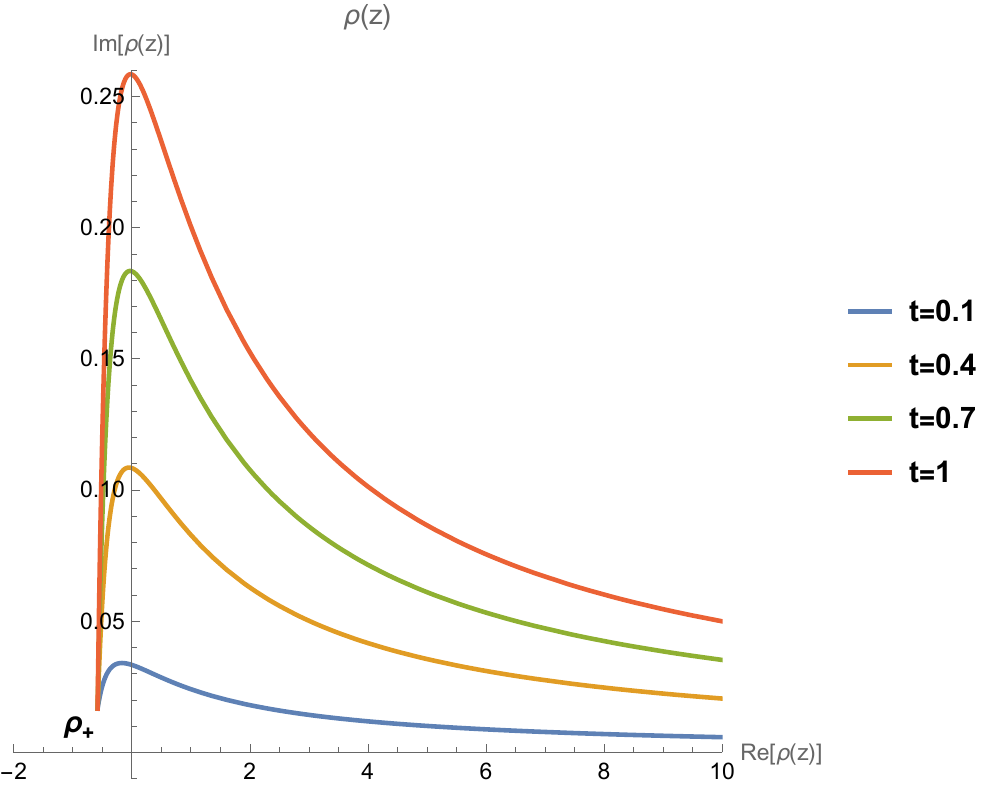}
\caption{The contour \eqref{eq:newcontour} for the $(+-)$ saddle with $\frac{2 \pi}{L}=0.05$ as an example. At large $L$, the saddle $\rho_+$ has a small imaginary part; the contour initially moves upward away from the singularity at the origin $\rho=0$, then converges to the positive real axis. Adjusting the value of ${\sf t}$ would allow us to control the boost away from the origin. Note the contour does not change the near-horizon and asymptotic Frobenius solutions as the transformation from $\rho$-domain to $z$-domain is still linear.}
\label{fig:newcontour}
\end{figure}
\FloatBarrier

Here we give a further remark about the choice of contour. For the $(+-)$ and $(--)$ saddles with negative real parts, numerically it is not a good idea to stick with the contour specified in \eqref{eq:contour} if we wish to probe large real values of $L$ within a reasonable number of grid points. The simple reason is that the singularity at $\rho=0$ would be on the right-hand side of the two saddles in the complex-$\rho$ plane, where the two saddles crossing the imaginary axis would be very close to it. One could of course introduce a new parameter to avoid this behavior (e.g. a contour with a homotopy parameter in \eqref{eq:contourt}), but it is still not an optimal choice if the contour converges much slower to the positive real axis (i.e. ${\sf t} \gg 1$ in \eqref{eq:contourt}). Instead, we find the following choice of contour has much better behavior
\be \label{eq:newcontour}
\rho(z)=(1-\sigma(z))(\rho_+ + \text{sgn}(\text{Im}[\rho_+]) \i {\sf t} \sigma(z))+ \sigma(z) |\Re[\rho_+]| \bigg(\frac{1+z}{1-z} \bigg),
\ee
with a function $\sigma(z)$ and a homotopy parameter ${\sf t}$. We pick $\sigma(z)=\frac{1+z}{2}$ such that the piece proportional to ${\sf t}$ would give an initial boost that moves away from the singularity at $\rho=0$, and after crossing the imaginary axis it approaches the positive real axis fast enough. This is best illustrated in Fig.~\ref{fig:newcontour}. In Section \ref{sec:othertopo}, we will consider scenarios with universes ending on $S^1 \times H^2$ and $S^1 \times S^1 \times S^1$ spatial topologies where the saddles would lie along the imaginary axis. These would be much closer to the singularity where we pick a similar contour.

\subsection{Rotational modes along the sphere}
\label{sec:rotaionalmodes}

We also expect another family of zero modes corresponding to the SO(3) rotations. The rotational modes are expected to be decoupled from the Schwarzian modes. We pick the following ansatz for them:
\be\label{eq:ROTANSQ}
h_{\mu \nu}\d x^\mu \d x^\nu=2 e^{\i k_n x}  \rho \sin^2{\theta} \d \phi \bigg(\i g_1 (\rho) \sqrt{f} \d x+g_2 (\rho) \frac{\d \rho}{\sqrt{f}} \bigg),
\ee
where $g_i$ with $ i \in \{1, 2 \}$ are functions of $\rho$ that are generally complex. This ansatz is adapted to the Killing vector of $S^2$ corresponding to rotations $\phi \to \phi+ \text{const}$. There are two other families of modes with ansatz adapted to the other two Killing vectors of $S^2$. The SO(3) symmetry implies that the eigenvalue spectrum should be the same regardless of the choice of Killing vector, and therefore we only consider the simplest one \eqref{eq:ROTANSQ}.

The traceless condition $h=0$ is automatically satisfied and the harmonic gauge 
\be
\nabla^\mu h_{\mu \nu}=0,
\ee
allows us to first solve
\be
g_1=-L\frac{(\rho f'+6 f)g_2+2 \rho f g'_2}{4 n \pi \rho},
\ee
then one can systematically reduce the spin-2 Lichnerowicz operator equation into a second-order ODE for $g_2(\rho)$, exactly parallel to the analysis we did in Section \ref{sec:SchwarzianSector}. We will adopt the collocation method again to study the generalized eigenvalue problem for $g_2$.

\smallskip
\begin{figure}[t!]
  \centering
  \begin{subfigure}[b]{0.6\textwidth}
    \centering
    \includegraphics[width=\textwidth]{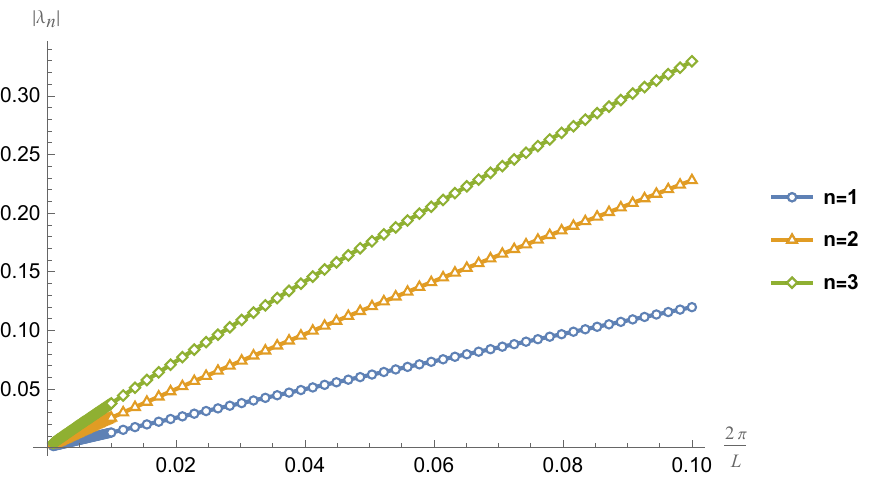}
    \caption{a: The eigenvalue spectrum over an extended range of $1/L$.}
    \label{fig:t10rotfullfit}
  \end{subfigure}
  
  \vspace{0.5cm}

  \begin{subfigure}[b]{0.45\textwidth}
    \centering
    \includegraphics[width=\textwidth]{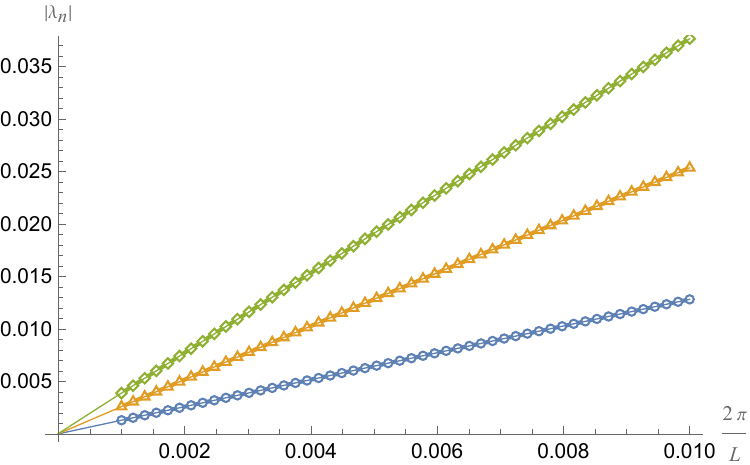}
    \caption{b: The eigenvalue spectrum at small $1/L$.}
    \label{fig:t10rotfit_zoomin}
  \end{subfigure}
  \hfill
  \begin{subfigure}[b]{0.45\textwidth}
    \centering
    \includegraphics[width=\textwidth]{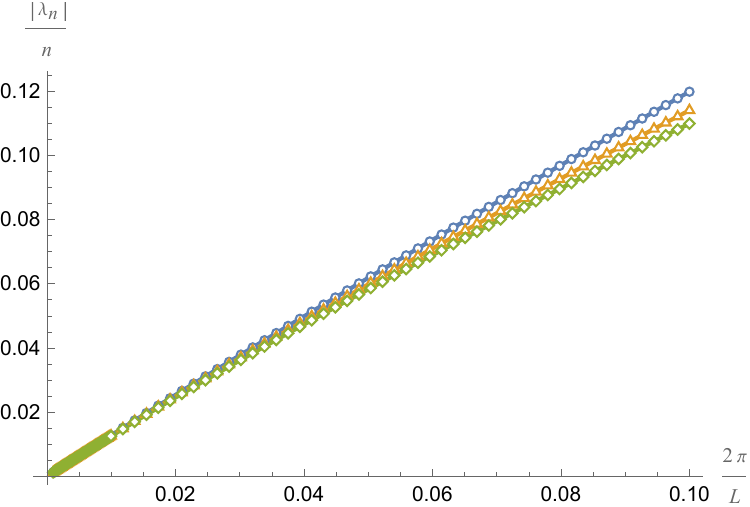}
    \caption{c: Rescaled eigenvalues.}
    \label{fig:t10rotrescale}
  \end{subfigure}
  
  \caption{The eigenvalue spectrum of the rotational modes by uplifting to the full Schwarzschild-dS geometry, along with the fit \eqref{eq:quadraticrotfit} demonstrating the linear behavior at small values of $1/L$. Here we sample $50$ points for $\frac{2 \pi}{L} \in [0.001,0.01]$ and $50$ points for $\frac{2 \pi}{L} \in [0.01,0.1]$. In Fig.~\ref{fig:t10rotrescale}, we similarly rescale with $\frac{|\lambda_n|}{n}$ to demonstrate $|\lambda_n| \sim k_n$.}
  \label{fig:t10roteigvalues}
\end{figure}

Now going to the compact $z$-coordinate with \eqref{eq:contour}, we perform a local Frobenius analysis. For the asymptotic boundary with $z=1$, the indicial equation gives
\be
s^2-5s-4 \lambda+4=0 \implies s_\pm=\frac{1}{2}(5 \pm \sqrt{9+16 \lambda}).
\ee
Near the horizon with $z=-1$, we have
\be
-4s'^2-4s'+n^2-1=0\implies s'_\pm=\frac{1}{2}(\pm n -1).
\ee
As explained in Section \ref{sec:SchwarzianSector}, the behavior at the two endpoints are consistent with the near-horizon behaviors of AdS black holes \cite{Kolanowski:2024zrq} with our choice of contour \eqref{eq:contour}. We motivate the following substitution
\be
g_2(z)=(1+z)^{\frac{1}{2}(|n|+1)}(1-z)^{\frac{3}{2}}w(z),
\ee
and restrict to $|n| \geq 1$. Notice that now it is only the zero-mode $n=0$ that is excluded. The eigenvalues from the rotational modes of the lowest three values of $n$ are depicted in Fig.~\ref{fig:t10roteigvalues}. Again we consider the following quadratic fit for $k_n/n=2\pi/L \in [0.001,0.01]$
\be \label{eq:quadraticrotfit}
|\lambda_n|=a_n+b_n\, k_n+c_n \, k_n^2 + \mathcal{O}(k_n^3),
\ee
where we find 
\be
a_1 \sim \mathcal{O}(10^{-6}), \quad b_1 \approx 1.297, \quad c_1 \approx -1.589.
\ee
\be
a_2\sim \mathcal{O}(10^{-5}), \quad b_2 \approx 1.293, \quad c_2 \approx -1.352.
\ee
\be
a_3\sim \mathcal{O}(10^{-5}), \quad b_3 \approx 1.290, \quad c_3 \approx -1.199.
\ee
Indeed it is true that the modes are becoming zero modes as we approach extremality at large $L$, since $k_n\propto 1/L$. We again verify that $b_{n}$ is approximately $n$-independent implying that the eigenvalue is proportional to $k_n\propto n$.

\smallskip
\begin{figure}[t!]
  \centering

\begin{subfigure}[b]{0.45\textwidth}
    \centering
    \includegraphics[width=\textwidth]{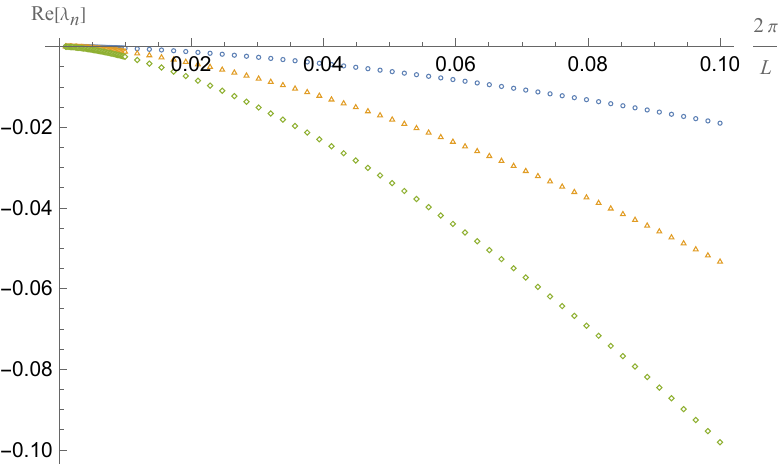}
    \caption{a: The real part of $\lambda_n$.}
    \label{fig:t10rotRe}
  \end{subfigure}
  \hfill
  \begin{subfigure}[b]{0.45\textwidth}
    \centering
    \includegraphics[width=\textwidth]{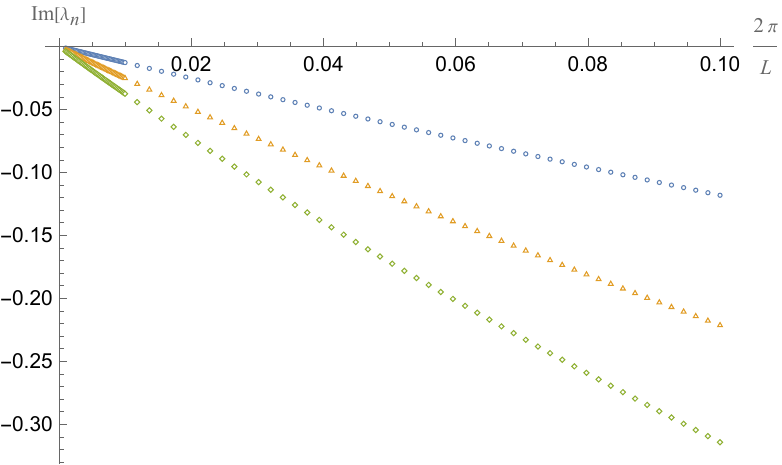}
    \caption{b: The imaginary part of $\lambda_n$.}
    \label{fig:t10rotIm}
  \end{subfigure}

    \vspace{0.5cm}

    \begin{subfigure}[b]{0.6\textwidth}
    \centering
    \includegraphics[width=\textwidth]{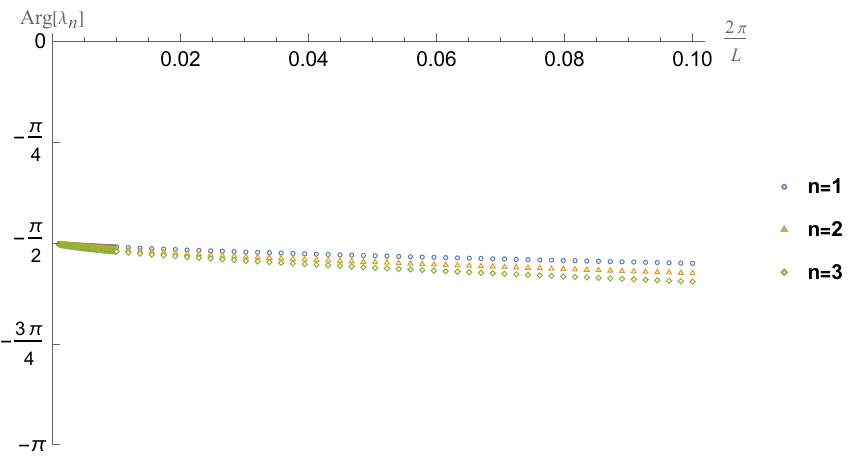}
    \caption{c: The phase of $\lambda_n$.}
    \label{fig:t10rotphase}
  \end{subfigure}
  
  \caption{In the top panel, we plot the real and imaginary parts of the eigenvalues $\lambda_n$ from the rotational modes for the three momenta modes over the extended range of $1/L$. In the figure below we plot the phase of $\lambda_n$, indicating that the eigenvalues are dominated by the imaginary part at large $L$.}
  \label{fig:t10roteigen}
\end{figure}

Again, the eigenvalues $\lambda_n$ are generally complex, as shown in Fig.~\ref{fig:t10roteigen}. But from reading the phase in Fig.~\ref{fig:t10rotphase} that we see $\text{Arg}[\lambda_n] \approx -\frac{\pi}{2}$ across the whole parameter regime of the three momenta modes, the eigenvalues at large $L$ are imaginary. This is indeed what we expect from a detailed analysis on the wavefunction in the large $L$ limit \eqref{eq:ONSAS1S2OMEGA} based on Kerr-dS geometry in Section \ref{sec:wavefunctionspin}, as we explain below.

\smallskip

\begin{figure}[t!]
\centering
\includegraphics[width=0.4\textwidth]{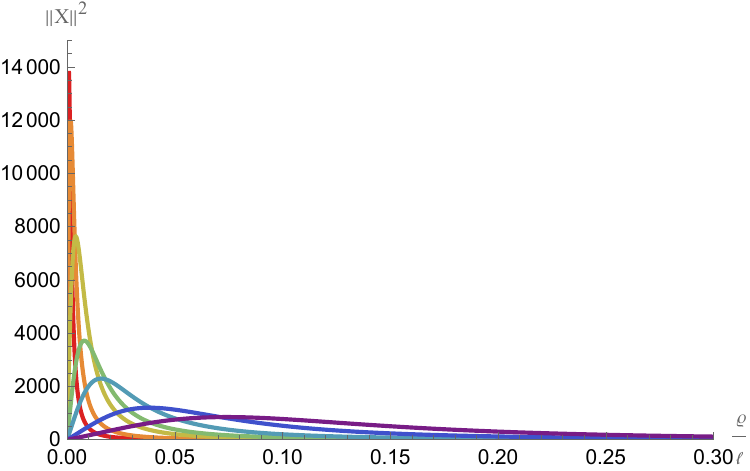}
\includegraphics[width=0.4\textwidth]{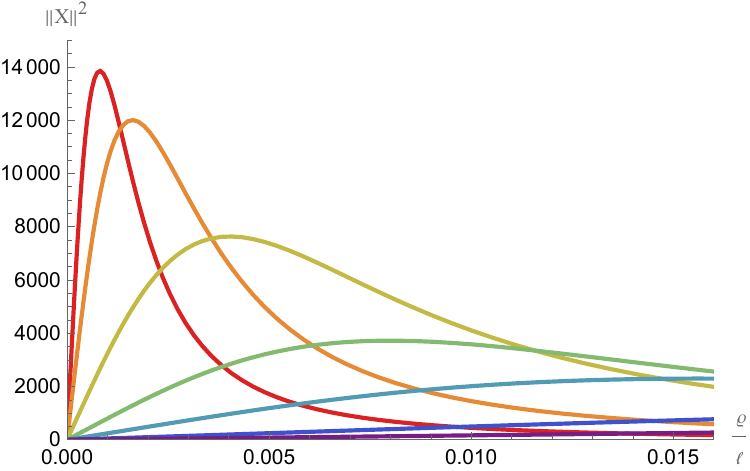}
\caption{The norm of the rotational modes defined as the absolute value of the integrand in \eqref{eq:norm_rot} for $n=2$ with $\frac{2 \pi}{L} \in [0.1, 0.05,  0.02, 0.01, 0.005, 0.002, 0.001]$. Again, we see that as we lower $1/L$, the norm is getting more and more localized toward the horizon, and the peak amplitude increases.} 
\label{fig:normrot}
\end{figure}

We similarly compute the eigenfunctions for the norm, we have
\bea \label{eq:norm_rot}
||h||^2&=&\int \d^4 x \sqrt{-g} \tilde{h}^{\mu \nu}h_{\mu \nu}
\no\\
&=&\int_{\rho_+}^\infty \d \rho \frac{16 \pi L \rho^2}{3} (g_2^2-g_1^2),
\eea
where we plot again the absolute value of the integrand above against the proper time, see Fig~\ref{fig:normrot}.

The implication of this section, similar to the previous one, is the existence of a local $\SU(2)$ set of light modes with action proportional to $S_{\text{dS}} k_n$. The contribution to the wavefunction, without keeping track of numerical prefactors and phases, is
\beq
\Psi_{\text{quantum}} \supset \frac{1}{\prod_{n\neq0} ( S_{\text{dS}} k_n)^{1/2} } \sim \frac{1}{S_{\text{dS}}^{1/2} L^{1/2}}.
\eeq
Recalling that there are two other families of modes obtained by rotations along $S^2$, the final contribution from the rotational modes is
\beq
\Psi_{\text{quantum}} \sim \frac{1}{S_{\text{dS}}^{3/2} L^{3/2}}.
\eeq
We emphasize that although the final power of $L$ is the same as the Schwarzian mode, the details in the origin of this effect are different. The rotational modes can be captured by an effective theory with action
\beq
\i S = \Phi_r  \int \d x\,  \{ F(x), x\} + K \int \d x\, \text{Tr}( g^{-1} \partial_x g)^2, ~~~~\Phi_r = \i \frac{\mu_N S_{\text{dS}}}{2\pi},~~~~K =- \i \frac{\mu_N S_{\text{dS}}}{4\pi}
\eeq
where $g(x) \in \SU(2)$. Each Fourier mode of $g(x)$ with a fixed momenta $k_n$ is in one-to-one correspondence with the modes in \eqref{eq:ROTANSQ}. Again, the precise coefficient of the action which we denoted by $K$ here can be determined by comparing with the classical action in the large $L$ limit. In this case we need a generalization of our analysis to incorporate rotation, which we do in Section \ref{sec:wavefunctionspin}, see in particular equation \eqref{eq:ONSAS1S2OMEGA}. The fact that $K$ is purely imaginary explains why we found modes that become purely imaginary in the large $L$ limit.

\subsection{Homotopy vs homology}
\label{sec:Homotopy}

As we explained in Section \ref{sec:REVIEW}, the choice of the background geometry involves a choice of contour in the $\rho$-plane that goes from the ``horizon'' $\rho_+$ towards the future Lorentzian time $\rho \to \infty$. If $\rho_+$ and $\mu$ were real, there would be a natural contour to take along the real axis such that the metric is real, but that is not the case. And even if that was the case, the question still remains on how are we supposed to treat saddles constructed by deforming the contour. Different contours are genuinely different geometries but we don't expect the need to perform a ``path integral'' over these deformations as part of the original gravitational path integral. This question was concretely raised by Witten recently in \cite{Witten:2021nzp} based on recent work by Kontsevich and Segal \cite{Kontsevich:2021dmb}.

\smallskip

We will address this question here in the context of our $S^1 \times S^2$ universe. What would happen if we integrated over all possible paths from $\rho=\rho_+$ to $\rho \to +\infty$? The first observation is that the on-shell action is independent of such a path as long as they keep the same end-points and are in the same homotopy class. Therefore, at the classical level, an integral over path deformations
\beq
\mathcal{C} = \{\text{Paths from $\rho_+$ to $+\infty$}\},
\eeq
would be badly divergent and seemingly redundant. A natural resolution is that we should sum over \textit{equivalence classes of contours} defined modulo homotopy
\beq
\mathcal{C}' = \{\text{Paths from $\rho_+$ to $+\infty$ modulo homotopy}\}.
\eeq 
For this prescription to make sense, not only the classical action should be invariant under homotopy but also quantum corrections. We will verify here in this section that the spectrum of $\Delta_L$, controlling leading-order quantum corrections to the wavefunction, are indeed invariant under homotopic deformations of the contour. This property is crucial in order to allow us to restrict the sum over $\mathcal{C}$ in the gravitational path integral to a sum over $\mathcal{C}'$ which can potentially be finite and well-defined. Since our problem can be ultimately reduced to an ODE, we will first give a general derivation that this is the case and then verify this numerically.

\smallskip

\begin{figure}[t!]
\centering
\includegraphics[width=0.6\textwidth]{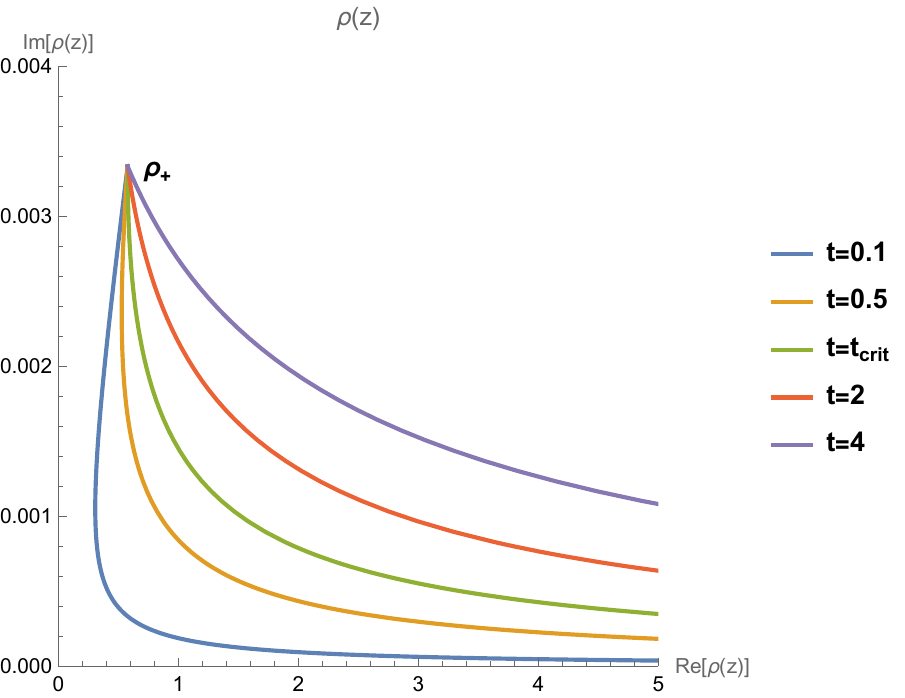}
\caption{The behaviors of the contour \eqref{eq:contourt} with different values of the homotopy parameter ${\sf t}$ for the physical saddle \eqref{eq:rhoplussaddle} with $\frac{2 \pi}{L}=0.01$. Here ${\sf t}_{\text{crit}}$ is the value of ${\sf t}$ where for ${\sf t} \geq {\sf t}_{\text{crit}}$ the contour is monotonic along the positive real axis. At large $L$, ${\sf t}_{\text{crit}} \gtrsim 1$. We see for larger values of ${\sf t}$, the contour would converge slower toward the positive real axis; while for small values of ${\sf t}$, the contour is not monotonic and would move "backward" first. Both extremes would require a much higher number of grid points in the numerical analysis.} 
\label{fig:homotopycontour}
\end{figure}

We have picked a simple choice of contour \eqref{eq:contour} in the analysis of the Schwarzian and rotational modes. It is easy to deform the contour by introducing a dimensionless homotopy parameter ${\sf t}$ in the design of the contour, i.e.
\be \label{eq:contourt}
\rho=\frac{1}{2}\rho_+ (1-z)+{\sf t}|\Re[\rho_+]| \bigg(\frac{1+z}{1-z} \bigg),
\ee
where in Fig.~\ref{fig:homotopycontour} we plot the behaviors of different contours for the physical saddle. We call them homotopy contours as they are continuously deformable to each other. In fact, one could consider a linear combination of different choices of contours that are homotopic as long as we keep the same boundary endpoints. The question is whether geometries corresponding to different values of ${\sf t}$ should be included in the gravitational path integral, or whether all of them should be identified as corresponding to the same unique saddle point. As explained in the previous paragraph, we will test this by showing that $\Delta_L$ is isospectral under changes in ${\sf t}$, which is a necessary condition in order to implement the identification of homotopy equivalent solutions. 

\smallskip

\begin{figure}[t!]
\centering
\includegraphics[width=0.6\textwidth]{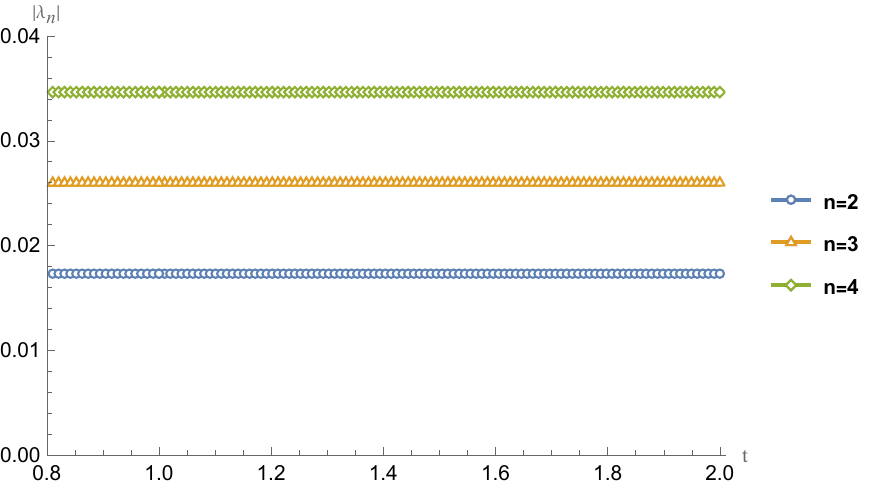}
\includegraphics[width=0.4\textwidth]{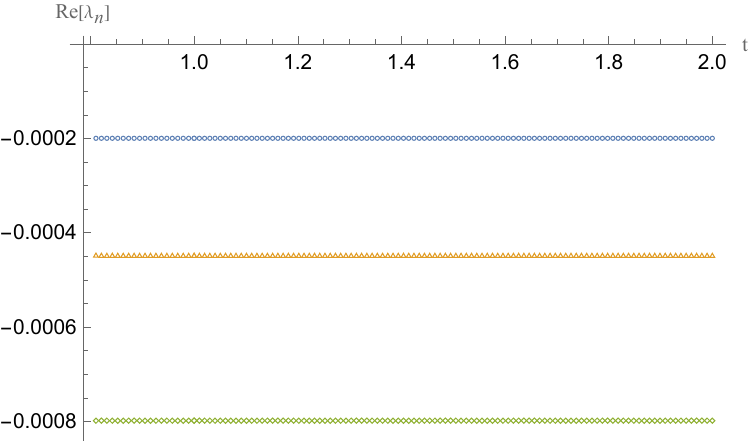}
\includegraphics[width=0.4\textwidth]{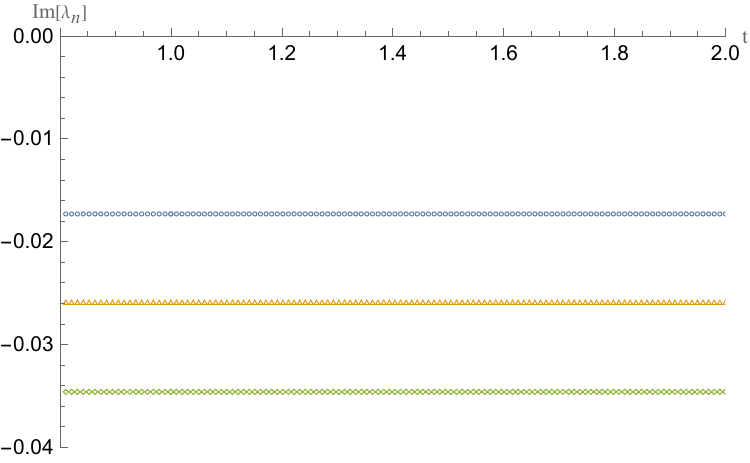}
\caption{We sample the eigenvalues corresponding to the Schwarzian modes by fixing $\frac{2 \pi }{L}=0.01$ but vary the homotopy parameter ${\sf t} \in [0.8, 2.0]$ for the lowest three momenta $n=2,3,4$. By maintaining a reasonable number of grid points, we found the eigenvalues are independent of ${\sf t}$ regardless of the contours being monotonic or not.}
\label{fig:homotopyfixL}
\end{figure}

\begin{figure}[t!]
\centering
\includegraphics[width=0.6\textwidth]{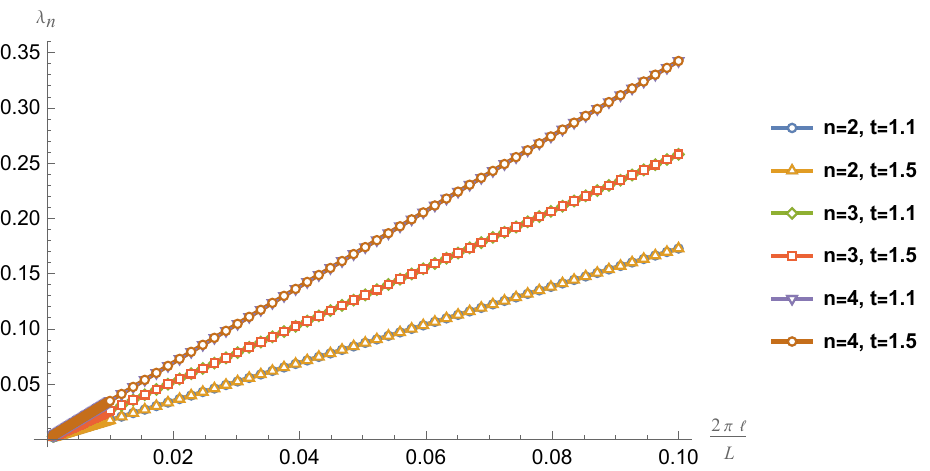}
\includegraphics[width=0.4\textwidth]{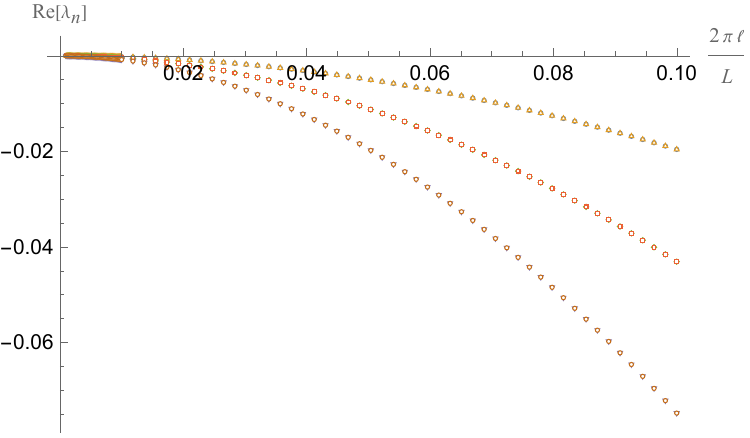}
\includegraphics[width=0.4\textwidth]{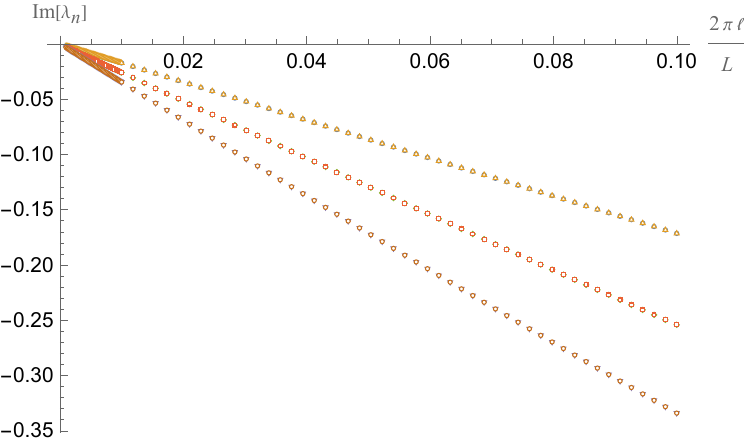}
\caption{We sample two different values of the homotopy parameter ${\sf t}$ but vary $1/L$ for the lowest three momenta $n=2,3,4$. It is clear that the eigenvalues overlap exactly. They also agree perfectly with Fig.~\ref{fig:t10eigvalues} and Fig.~\ref{fig:t10eigen} where ${\sf t}=1$.}
\label{fig:homotopyfixt}
\end{figure}

As we have premiered when we describe a different contour choice in \eqref{eq:newcontour}, one may have noticed that a contour change does not affect the spectrum of the Schwarzian modes. In Fig.~\ref{fig:homotopyfixL} and Fig.~\ref{fig:homotopyfixt}, we present further numerical evidence that the spectrum of the Schwarzian modes are independent of the homotopy parameter ${\sf t}$.\footnote{While for other $L$-independent eigenvalues, one needs higher grid points to show that they are also independent of ${\sf t}$.}

\smallskip

Different contours are exactly isospectral. But what would be the criteria? The answer is simple in the context of the second-order ODE at hand and its corresponding differential operator $\hat{D}(\rho)u(\rho)=0$. Two ODEs are isospectral if they are related by a similarity transformation when we perform the map that also depends on the homotopy parameter: $\{f_{{\sf t}}(z): \rho \to \rho(z)\}$ with $z \in [-1, 1]$, where then $\hat{D}_{{\sf t}}(z) U(z)=0$ with $\hat{D}_{{\sf t}}=S_{{\sf t}} \hat{D} S_{{\sf t}}^{-1}$. If we could find such an operator $S_{{\sf t}}(z)$, then we can prove that different contours defined by the homotopy parameter ${\sf t}$ are also isospectral since 
\be
\hat{D}_{{\sf t}_1}=S_{{\sf t}_1}\hat{D}S^{-1}_{{\sf t}_1}=S_{{\sf t}_1}(S^{-1}_{{\sf t}_2}\hat{D}_{{\sf t}_2}S_{{\sf t}_2})S^{-1}_{{\sf t}_1}= (S_{{\sf t}_1}S^{-1}_{{\sf t}_2})\hat{D}_{{\sf t}_2}(S_{{\sf t}_2}S^{-1}_{{\sf t}_1}).
\ee
Hence $\hat{D}_{{\sf t}_1}$ are indeed isospectral to $\hat{D}_{{\sf t}_2}$ related by a similarity transformation. In fact, we simply pick $S_{\sf t}$ to be the composition operator such that it composes the function in the $\rho$-domain to the $z$-domain
\be
(S_{{\sf t}}\,u)(z)=u \circ f_{\sf t}(z)=u(\rho)|_{\rho \to f_{\sf t}(z)},
\ee
where we note it is not an identity operator unless the map is $f_{\sf t}(z)=z$. Then the only requirement is that the inverse $S_{\sf t}^{-1}$ exists provided the map is globally invertible $z=f^{-1}_{\sf t}(\rho)$ converting the function from $z$-domain back to the $\rho$-domain
\be
(S_{\sf t}^{-1}U)(\rho)=U \circ f^{-1}_{\sf t}(\rho)= U(f_{\sf t}^{-1}(\rho))=u(\rho).
\ee
Of course, the map must not change the endpoints in any singular ways and the contours do not cross any branch cuts or essential singularities. Hence we have the same boundary value problem. One can explicitly check the similarity transformation will correctly generate the same ODE in the $z$-domain that is derived by merely performing the map directly.

\smallskip

When is the map $f_t(z)$ globally invertible? If all the parameterizations are real numbers $\mathbb{R}$, then it is required that the map is a monotonic function for such a bijective map. However, now we have complex saddles where we map a contour in the complex plane to a real interval $z \in [-1, 1]$, then monotonicity is only sufficient but not necessary. In fact, this can be numerically confirmed in Fig.~\ref{fig:homotopyfixL} for homotopy parameter ${\sf t}< {\sf t}_{\text{crit}}$ where the contour is not monotonic along the positive real axis.

\smallskip

It is precisely the requirement of global invertibility that allows us to claim that the gravitational path integrals will be equal at the one-loop level for all contours related by homotopy. We believe this holds at higher-loop levels, although we will not attempt to give a general proof. \emph{This  justifies replacing the sum over saddles, as characterized by all contours, by a sum over equivalence classes of saddles, defined modulo homotopy. }

\smallskip

As discussed in \cite{Witten:2021nzp}, a non-trivial sum over "homology" contours remains since $\mathcal{C}'$ is not necessarily trivial. By a homology contour, we mean that we fix the same boundary endpoints and the boundary conditions, but we allow the contour to intersect itself in the complex $\rho$-domain. Then the map is no longer injective and cannot be globally invertible. It is possible that locally if we only study a set of eigenvalue branches like the Schwarzian modes, homology contours may give identical results. From the gravitational path integral they should be considered as distinct saddles.

\smallskip

Finally, let us mention that the same analysis applies to the original Hartle-Hawking saddle for an $S^3$ universe, see discussion in Section \ref{sec:REVIEW}. Our arguments show that the gravitational path integral, even at the quantum level, is independent of the path taken between $t=\i \pi/2$ and $t\to\infty$ as long as they are related by homotopy. The contour shown in Fig.~\ref{fig:HHS3} is the original one proposed by Hartle and Hawking, corresponding to the half-sphere glued to Lorentzian global de Sitter. Another important contour is the "-AdS" one, which runs along the horizontal line $\text{Im}(t)=\pi/2$ and eventually merges the real axis, see \cite{Maldacena:2002vr,Hertog:2011ky, Harlow:2011ke,Maldacena:2019cbz}. The path integral over both contours will be the same at the quantum level.

\smallskip

\subsection{Assembling the pieces}\label{sec:assembling}

The final result for the wavefunction, in the large $L$ limit and at the quantum level, is
\beq
\Psi(S_L^1 \times S^2) =  \frac{c_1 S_{\text{dS}}^{c_2}}{L^{3}} \, \exp{\left(\frac{S_{\text{dS}}}{3}-\frac{8\pi^2 S_{\text{dS}}}{27 L^2}\right)} \, \exp{\left(\i S_{\text{div}}-\i \frac{\mu_NS_{\text{dS}}}{2 \pi}L+ \ldots\right)},
\eeq
where the dots in the second exponential are purely imaginary and subleading in the large $L$ limit. A power of $L^{-3/2}$ arises from the Schwarzian modes. The rotational mode comes in a triplet, each giving rise to a factor of $L^{-1/2}$ and producing a final factor of $L^{-3/2}$ which combines with the Schwarzian modes to produce $L^{-3}$. Computing the $L$-independent and $S_{\text{dS}}$ independent coefficients $c_1$ and $c_2$ is beyond the scope of this paper. In particular, we expect $c_2$ to be a real number while $c_1$ could potentially be complex,
\beq
c_1 \in \mathbb{C},~~~c_2 \in \mathbb{R},
\eeq
as discussed at the beginning of this section.

\smallskip

It will be useful, for later considerations, to elaborate on the analysis in \cite{Maldacena:2019cbz}. After incorporating the quantum correction, what is the most probable relative size $L$ of the circle? To answer this question we define the (un-normalized) probability distribution
\bea
P(S_L^1 \times S^2) &=& \frac{1}{L} \, |\Psi(S_L^1 \times S^2)|^2 ,\nonumber\\
&\sim& \frac{|c_1|^2 S_{\text{dS}}^{2c_2}}{L^{7}} \, \exp{\left(\frac{2S_{\text{dS}}}{3}-\frac{16\pi^2 S_{\text{dS}}}{27 L^2}\right)}.
\ea
An extra factor of $L$ in the definition of the probability distribution, as related to the wavefunction, comes from a residual translational gauge symmetry, and it is justified in Appendix G of \cite{Maldacena:2019cbz}, see \cite{Cotler:2024xzz} for a discussion in the context of JT gravity. The probability distribution is now normalizable at large $L$. The Schwarzian modes ensure that the answer remains finite; otherwise, even with the factor of $1/L$, it would be logarithmically divergent. Moreover, by solving $\partial_L \log P |_{L=L_0} = 0$ we can find the most likely value of this parameter
\beq
L_0 \sim \frac{4\sqrt{2}\pi}{3\sqrt{21}} \sqrt{S_{\text{dS}}},~~~~~~P(L_0) \sim S_{\text{dS}}^{2c_2 -\frac{7}{2}} \,e^{\frac{2 S_{\text{dS}}}{3}},
\eeq
to leading order in $S_{\text{dS}}\gg 1$. Notice that $L_0 \sim \sqrt{S_{\text{dS}}} \ll S_{\text{dS}}$. This means that the most likely universe is one in which the semiclassical Schwarzian and $\SU(2)$ rotational mode analysis is valid. There are no higher-order corrections in the Schwarzian sector beyond one-loop, but if $L \sim S_{\text{dS}}$ there would be non-perturbative corrections arising from the rotational mode. See Appendix \ref{app:SU2} for some comments on this. Since the most likely size $L_0$ is much smaller than the de Sitter entropy, we do not need to worry about this issue.

\smallskip

The Schwarzian quantum corrections are crucial to get a sensible answer. As already pointed out, they render the probability of creating an $S^1_L \times S^2$ universe finite after integrating over $L$. An infinite answer would mean that creating an $S^1 \times S^2$ universe is more likely than $S^3$, which has a finite norm. Moreover, at the classical level, the final expression for the probability distribution matches with the path integral on the exact Nariai spacetime $S^2\times S^2$, which solves the classical field equations  when the spheres take an appropriate radius, see \cite{Ginsparg:1982rs}. In other words, at the exponential level, the probability is equal to the $S^2\times S^2$ on-shell action
\beq
P(L_0) \sim e^{\i S[S^2 \times S^2]},~~~~\i S[S^2 \times S^2] = \frac{2 S_{\text{dS}}}{3}.
\eeq
Of course extending this correspondence to the quantum level is an open problem, particularly reproducing the overall powers of $S_{\text{dS}}$, as well as the overall phase. This analysis, in particular with respect to the evaluation of the phase of the $S^2\times S^2$ path integral, will appear in a separate publication \cite{WOP_ST}.

\section{Generalizations}

In this section we will first consider the generalization to universes with different topologies than $S^3$ or $S^1\times S^2$. We will then consider how the wavefunction of the $S^1 \times S^2$ universe depends on the choice of spin structure on a spatial slice.  
\subsection{Other spatial topologies}
\label{sec:othertopo}

We first generalize the calculations done earlier to other spatial topologies. We begin by considering an $S^1 \times \Sigma_g$ universe. The classical metric that implements the no-boundary prescription for the wavefunction of the universe is 
\beq
\d s^2 = - \frac{\d \rho^2}{f} + f \d x^2 + \rho^2 \d \Sigma_g^2 ,~~~~~f = \rho^2 + \frac{\mu}{\rho} + 1,
\eeq
where $\Sigma_g$ is a compact hyperbolic surfaces of genus $g$ and constant negative curvature $R_\Sigma=-2$. The latter condition is required to solve 4d Einstein's equations. The future asymptotics of the solution is
\beq
\d s^2 \sim -\frac{ \d \rho^2}{\rho^2} + \rho^2( \d x^2 +  \d \Sigma_g^2),~~~~\rho\to\infty.
\eeq
The metric can contract smoothly at the location $\rho_+$ related to the length of the circle by
\beq
L = \pm \i \frac{4\pi}{ f'(\rho_+)} = \pm \i \frac{4\pi \rho_+}{1 + 3 \rho_+^2}.
\eeq
There are four solutions to this equation
\beq
\rho_+ = \pm \frac{2\pi \i}{3L} \pm \frac{\i \sqrt{3L^2 + 4\pi^2}}{3L},
\eeq
Notice that they are all pure imaginary, regardless of $L$. The reason is that there is no Lorentzian Nariai limit with hyperbolic topology. Nevertheless, this does not imply one cannot find a complex no-boundary geometry evaluating the wavefunction. The purpose of this section is to analyze the physics implied by these complex metrics.

\smallskip

Which saddle out of the four shall we consider? We can immediately rule out two saddles according to the rule that the real part of the proper Euclidean time between $\rho_+$ and $\rho\to\infty$ should be positive or, equivalently, the imaginary part of Lorentzian proper time should be negative, namely
\beq\label{eq:S1H2RPT}
\text{Im}\left( \int_{\rho_+}^\infty \frac{\d \rho}{\sqrt{f(\rho)}}\right) <0.
\eeq
This condition is only satisfied by the following two saddles
\beq\label{eq:S1H2RP12}
\rho_+ = \frac{2\pi \i}{3L} \pm \frac{\i \sqrt{3L^2 + 4\pi^2}}{3L}.
\eeq
In the case of an $S^1 \times S^2$ universe, the two saddles that satisfied \eqref{eq:S1H2RPT} were in a sense equally good, and one needed further information to determine which one is physical. This would require a careful analysis of the path integral contour. In \cite{Maldacena:2019cbz}, it was argued that a contour rotation from reasonable Lorentzian metrics suggests including only one of the two saddles. 

\smallskip 

For a topology $S^1 \times H^2$, can we determine which of the two solutions \eqref{eq:S1H2RP12} is the physical one, or should we include both? The situation in this case is simpler. Compute the classical approximation to the wavefunction first for either saddle. We evaluated the wavefunction by parameterizing the proper length of $S^1$ by $L_{proper} = \L \sqrt{f(\rho_b)}  L$ and obtained for large $L_{proper}$, or equivalently large $\rho_b$, the result
\beq
\Psi(S^1_L \times \Sigma_g) \sim \text{exp}\left(\frac{(g-1)\L^2}{4G_N}\i (1-\rho_+^2)\rho_+ L-\underbrace{\i \frac{(g-1)\L^2}{G_N} L(\rho_b^3 +  \rho_b/2)}_{\text{pure phase}}  \right),
\eeq
where we used the Gauss-Bonnet theorem that says the volume of $\Sigma$ is $4\pi(g-1)$. The first term is finite in the large $\rho$ limit. Since we found for either saddle, $\rho_+$ is purely imaginary, the first term is purely real. The second term is a divergent phase that depends on the size of the universe in the future, parameterized by $\rho$. It is similarly given by an integral of local quantities in the future $S^1 \times H^2$. 

\smallskip 

Next we evaluate this answer for the two saddles in \eqref{eq:S1H2RP12}, in the limit that $L\to\infty$. For the saddle with the positive sign $\rho_+ = \frac{2\pi \i}{3L} + \frac{\i \sqrt{3L^2 + 4\pi^2}}{3L}$ we obtain a classical contribution
\beq
\Psi(S^1\times \Sigma_g) \sim \text{exp}\left(- \frac{(g-1) \L^2}{3\sqrt{3}G_N} L - \frac{\pi (g-1)\L^2}{3G_N} - \frac{2\pi^2 (g-1)\L^2 }{3 \sqrt{3} G_N L} + \ldots  \right),
\eeq
and for the saddle with the minus sign $\rho_+ = \frac{2\pi \i}{3L} -\frac{\i \sqrt{3L^2 + 4\pi^2}}{3L}$ we obtain
\beq
\Psi(S^1\times \Sigma_g) \sim \text{exp}\left(  \frac{(g-1) \L^2}{3\sqrt{3}G_N} L - \frac{\pi (g-1)\L^2}{3G_N} + \frac{2\pi^2 (g-1)\L^2 }{3 \sqrt{3} G_N L} + \ldots   \right),
\eeq
where the dots denote terms that are subleading in the large $L$ limit, as well as purely imaginary terms that only lead to phases of the wavefunction. We see, without needing to discuss the path integral contour, that only the plus sign in $\rho_+$ makes sense (the top equation). The other solution (bottom equation) is unphysical since it leads to an unnormalizable wavefunction. The probability distribution diverges as $|\Psi|^2 \sim e^{L}$ and we still need to integrate over the parameter $L$, which cannot be convergent. Moreover, there is a value of $L$ for which a universe with topology $S^1 \times H^2$ would be more likely to be created than an $S^3$ universe, which seems unphysical. The top solution has a convergent behavior as $L\to\infty$ (as well as $L\to0$). It also has a convergent behavior in the genus of the hyperbolic surface. For the bottom equation, the most dominant surface has $g\to\infty$ while for the top equation it is $g=2$. 

\smallskip 

The above considerations imply that the only possibly physical solution is 
\beq \label{eq:S1H2saddle}
\text{Physical:}~~~~\rho_+= \frac{2\pi \i}{3L} + \frac{\i \sqrt{3L^2 + 4\pi^2}}{3L}.
\eeq
The classical wavefunction of this solution is
\be
\Psi \sim \text{exp} \bigg[ -(g-1) \bigg( \frac{(3 \pi +\sqrt{3L^2+4 \pi^2})S_{\text{dS}}}{9 \pi}+\frac{4 \pi (2 \pi \sqrt{3 L^2+4 \pi^2}) S_{\text{dS}}}{27 L^2} \bigg)+\i (\ldots) \bigg].
\ee
In the limit of small circles $L\to 0$ it is clear that the real part of the action becomes arbitrarily large and negative as $-1/L^2$. This implies that the wave function vanishes fast for small $L$, but we also saw that it vanishes for large $L$. Unlike what happens for $S^1 \times S^2$, there is a classical value of $L$ that is finite and maximizes the probability $|\Psi|^2$. This value is $L= 2\pi$. For this size of $L$, the geometry is nothing else than AdS in the Rindler coordinates, and the probability is
\bea
\text{max}_L\,  |\Psi(S^1\times \Sigma_g)|^2 \sim  e^{- (g-1)S_{\text{dS}}},
\ea
while this takes the largest value for $g=2$, since it is the smallest genus that allows for a compact space.

\smallskip

The universe most likely created with a spatial $S^1 \times H^2$ has a finite $L$. The quantum effects we studied in Section \ref{sec:QuantumCorrections} are relevant when $L$ becomes large, which is in the tail of the probability distribution. However, it is worth exploring what happens to the one-loop determinant in this limit. We expect to find Schwarzian modes just as in Section \ref{sec:QuantumCorrections}. The only difference now is that we expect the Schwarzian coupling to be real instead of imaginary (the $1/L$ term in the action is real). 

\smallskip

\begin{figure}[hbt!]
\centering
\includegraphics[width=0.7\textwidth]{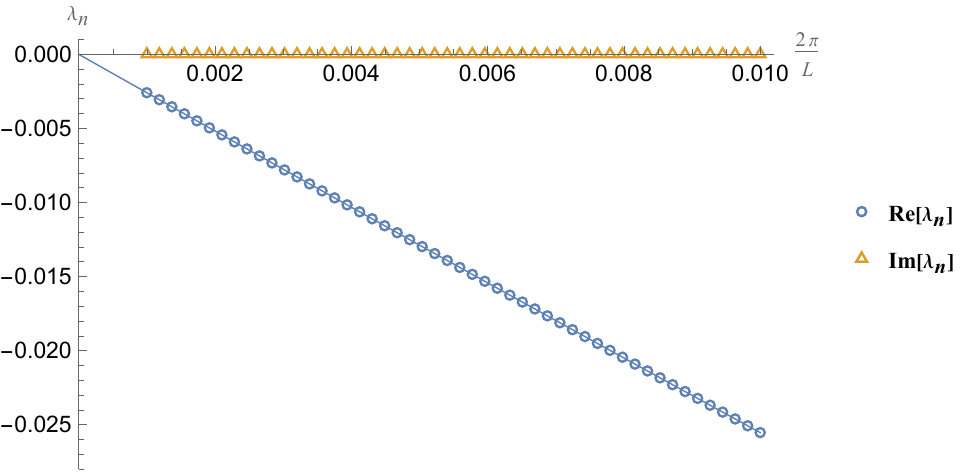}
\caption{Lowest eigenvalue for the $S^1 \times H^2$ universe with momentum $n=3$. The imaginary part is of order $O(10^{-20})$ and therefore negligible. The real part vanishes linearly, as expected.}
\label{fig:H2modes}
\end{figure}

This is indeed true. For the physical saddle, the Schwarzian modes do appear at large $L$ and the eigenvalues are real, see Fig.~\ref{fig:H2modes}. As we have commented near the end of Section \ref{sec:SchwarzianSector}, the physical saddle \eqref{eq:S1H2saddle} lies on the imaginary axis, hence it would be ideal to pick a contour similar to \eqref{eq:newcontour} by replacing $|\Re[\rho_+]|$ with $|\text{Im}[\rho_+]|$ in order to probe large values of $L$ within a reasonable number of grid points. Note the near-horizon and asymptotic Frobenius solutions are exactly the same as the $S^1 \times S^2$ case. The eigenvalue is negative but not large enough to spoil our ansatz, see discussion around \eqref{eq:lambdabound}. 

\smallskip

\FloatBarrier

\newpage

Finally, we analyze the case of a universe with topology $S^1 \times S^1\times S^1$. The metric is now 
\beq
\d s^2 = - \frac{\d \rho^2}{f} + f \d x^2 + \rho^2 (\d y^2 + \d z^2) ,~~~~~f = \rho^2 + \frac{\mu}{\rho}.
\eeq
We take the $y$ and $z$ coordinates to have period $L_y$ and $L_z$ respectively. The future spatial metric for large $\rho$ is given by
\beq
\d s^2 \sim - \frac{\d \rho^2}{\rho^2} + \rho^2 \Big(  \d x^2 + \d y^2 + \d z^2\Big). 
\eeq
The equation that determines $\rho_+$ now becomes much simpler and leads to only two solutions $\rho_+ = \pm \frac{4 \pi \i }{3 L}$, instead of four. We can verify that the solution with a minus sign $\rho_+ = - \frac{4 \pi \i }{3 L}$ leads to a proper time evolution with positive imaginary part namely $\text{Im}(\int_{\rho_+}^\infty \d\rho/\sqrt{f})>0$. This leaves us with only one possible physical solution which is
\beq \label{eq:T3saddle}
\rho_+ = \frac{4 \pi \i }{3 L}.
\eeq
The on-shell action in this case is very simple. It is given by
\beq
\Psi(S^1 \times S^1 \times S^1) \sim \text{exp}\left(-   \frac{4\pi L_y L_z}{27 L^2} S_{\text{dS}}- \i \frac{ LL_yL_z}{4\pi^2} S_{\text{dS}} \rho^3_b \right).
\eeq
We see that the imaginary term is now directly proportional to the volume of the future slice since the proper lengths of each circle are $\rho_b \L L$, $\rho_b \L L_y$ and $\rho_b \L L_z$ and the volume is simply the products of these factors. The first term is real and local in the $y$ and $z$ directions but not in $x$.

\smallskip

A consequence of the observations in the previous paragraph is that the absolute value of the wavefunction $|\Psi|^2$ is not invariant under the exchange of the three circles $(x,y,z)$, while the future spatial $S^1 \times S^1 \times S^1$ certainly is. The reason is obvious, when we wrote down the no-boundary geometry we singled out one circle, denoted by $x$, to smoothly contract in the past while the other two remain non-contractible. But we could have taken any of the three circles to contract which would lead to a different geometry. The symmetry under the exchange of the three circles, or more generally the modular invariance of the spatial $T^3$, has to be realized by a sum over geometries. This is a similar situation to the $\SL(2,\mathbb{Z})$ black holes that appear in $AdS_3$ with a time-like toroidal boundary.

\smallskip

\begin{figure}[hbt!]
\centering
\includegraphics[width=0.7\textwidth]{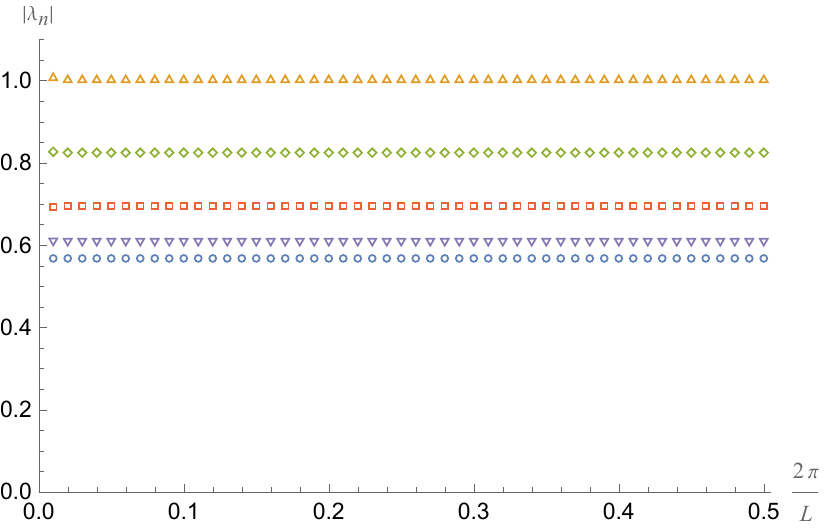}
\includegraphics[width=0.45\textwidth]{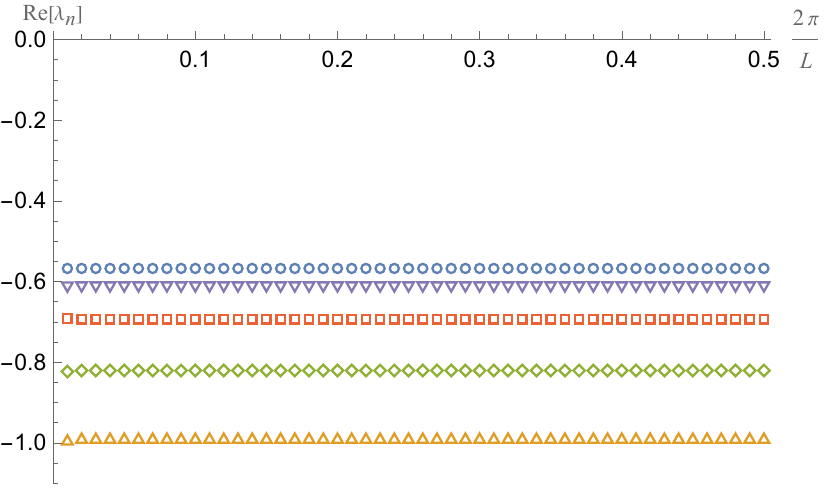}
\includegraphics[width=0.45\textwidth]{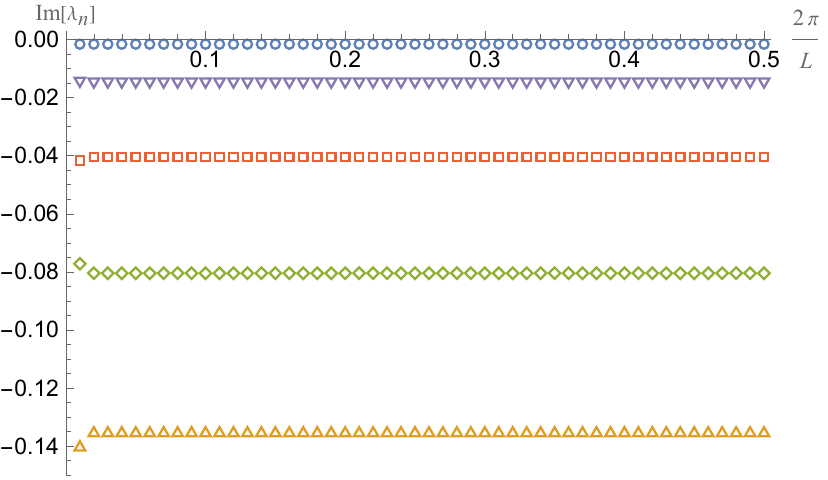}
\caption{Here we plot the lowest five eigenvalues for the $S^1 \times S^1 \times S^1$ with momentum $n=3$. Since for large $L$ we would be very close to the singularity $\rho=0$ given the physical saddle \eqref{eq:T3saddle}, we sample a much wider range of $\frac{2 \pi}{L} \in [0.01,0.5]$. By pushing to a sufficiently large number of grid points, we can see not only $|\lambda_n|$, but also their real and imaginary parts stabilize. We do not observe any $L$-dependent mode, this is a strong indication that there is no such mode.} 
\label{fig:T3modes}
\end{figure}

What is the fate of quantum corrections in the large $L$ limit for the toroidal universe? The answer above suggests that the Schwarzian coupling is zero, since now there is no linear term in $1/L$. Indeed, the Schwarzian modes just disappear and there do not seem to be any light mode emerging in the large $L$ limit. Again the contour we used is \eqref{eq:newcontour} by replacing $|\Re[\rho_+]|$ with $|\text{Im}[\rho_+]|$, and the near-horizon and asymptotic Frobenius solutions are exactly the same as the $S^1 \times S^2$ case. The result appears in Fig.~\ref{fig:T3modes}. It is important to notice that in this case the large $L$ limit is singular since $\rho_+$ approaches the origin. Nevertheless, it is still a reasonable question to ask whether Schwarzian-like modes are present at large but finite $L$.

\FloatBarrier

\subsection{Dependence of the wavefunction on the spin structure}
\label{sec:wavefunctionspin}

So far we have discussed the possibility of creating a universe with a round spatial $S^1 \times S^2$. If we include fermions in our theory, we can ask how this probability depends on a choice of spin structure. Since $S^2$ is simply connected, there is a unique choice of spin structure, but fermions could be periodic or anti-periodic along the spatial $S^1$. The goal of this section is to exploit recent developments \cite{Cabo-Bizet:2018ehj,Iliesiu:2021are,Chen:2023mbc} to give a quantitative answer to this question utilizing the gravitational path integral.

\smallskip

Before discussing the question of the spin structure, it is useful to consider a slightly more general setup. Consider a future spatial universe $S^1 \times S^2$ with a defect localized in $S^1$. The role of the defect is simply to impose the following identification on all fields which we collectively denote by $\Phi$, namely
\beq\label{eq:idtwde}
\Phi(x,\phi) = \pm \Phi(x+ L,\phi + \alpha),
\eeq
where the sign is positive for bosons and negative for fermions and $\phi$ generates a $\U(1)$ isometry of the $S^2$. As we go around the circle the defect implements a rotation on the two-sphere. We will evaluate the wavefunction of the universe for an arbitrary parameter $\alpha$. The case $\alpha=0$ corresponds to the removal of the defect and antiperiodic fermions along $S^1$. The case $\alpha = 2\pi$ is equivalent to removing the defect but imposing periodic boundary conditions for fermions.

\smallskip

In the presence of the defect, the Schwarszchild-dS metric we used so far does not work. The periodic identification is inconsistent with the circle that contracts far in the past. The saddle that allows us to smoothly implement the no-boundary proposal is the Kerr-dS black hole 
\be \label{eq:KerrdS}
\d s^2=-\frac{\Delta_\rho}{{\varrho}^2} \bigg[\d x-\frac{a \sin^2{\theta}}{\Xi} \d \tilde{\phi} \bigg]^2+\frac{\varrho^2}{\Delta_\rho}\d\rho^2+\frac{\varrho^2}{\Delta_\theta}\d \theta^2+\frac{\Delta_\theta \sin^2{\theta}}{\varrho^2}\bigg[a \d x-\frac{\rho^2+a^2}{\Xi} \d \tilde{\phi}\bigg]^2,
\ee
where
\be
\varrho^2=\rho^2+a^2 \cos^2{\theta} ,\quad \Xi=1+a^2,
\ee
\be
\Delta_\rho=(\rho^2+a^2)(1-\rho^2)-\mu \rho, \quad \Delta_\theta=1+a^2\cos^2{\theta}.
\ee
The geometry now depends on two parameters $\mu$ and $a$ which can be matched to $L$ and $\alpha$. The periodicity condition \eqref{eq:idtwde} will be imposed by identifying the cycle that contracts in the far past. The angle $\tilde{\phi}$ is related to the angle $\phi$ that appears in \eqref{eq:idtwde} via $\phi = \tilde{\phi} - a x$. We wrote the metric in terms of $\tilde{\phi}$ simply to follow the conventional presentation of the Kerr-dS metric. 

\smallskip

In Lorentzian signature, the solution is interpreted as a black hole and has horizons at the locations where $\Delta_\rho(\rho)=0$. This equation has four solutions, for a given $\mu$ and $a$, or in other words for a given mass and angular momentum of the black hole. Two of these roots correspond to the outer and inner black hole horizons for the rotating solution. The third root corresponds to the cosmological horizon, while the fourth is unphysical in Lorentzian signature. For recent work on a careful account of the different extremal limits of this geometry see \cite{Castro:2022cuo}.

\smallskip

We will be interested in this geometry in the context of the no-boundary wavefunction proposal. In this case we are interested in geometries that end at $\rho \to +\infty$ and evolve towards the past in the complex plane until they reach a root $\rho_+$ consistent with the choice of $L$ and $\alpha$, similar to what we did in Section \ref{sec:REVIEW}. In this context it is useful to eliminate the mass parameter in favor of $\rho_+$ via $\mu=(\rho^2_++a^2)(1-\rho^2_+)/\rho_+$. Demanding that the horizon is smooth we obtain the following equations relating $\rho_+$ and $a$ to $L$ and $\alpha$, namely
\beq
L= \pm \frac{4\pi \i(\rho^2_++a^2)}{\L\rho_+\bigg(1-a^2-3 \rho_+^2-\frac{a^2}{\rho_+^2}\bigg)}, ~~~~~ \frac{\alpha}{ L}=\frac{a (1-\rho^2_+)}{\rho^2_++a^2}.
\eeq
To obtain these equations, we can work in dS signature, or begin with an analytic continuation of the Kerr-AdS metric, evaluate the inverse temperature $\beta$ and the angular velocity $\Omega$, and then analytically continue to dS. The length of the thermal circle is given by $\i \beta = \pm L$ while the twist angle is $\alpha = \i \beta \Omega $. Using the explicit expressions for Kerr-AdS we obtain the results quoted above.

\smallskip

We want to compute the wavefunction of the universe $\Psi$ with geometries ending on $S^1 \times S^2$ in the expanding region. We will consider large enough $\rho$ such that we are in an expanding region and then $\rho$ becomes timelike with $\Delta_\rho<0$. As $\rho \to \infty$, the metric is approximately
\be
\d s^2 \approx \rho^2\d x^2-\frac{2 a \rho^2 \sin^2{\theta}}{1+a^2}\d x \d\tilde{\phi}-\frac{\d \rho^2}{\rho^2} +\frac{\rho^2}{\Delta_\theta}d \theta^2+\frac{\rho^2 \sin^2{\theta}}{\Xi} \d \tilde{\phi}^2,
\ee
and we define \cite{Hawking:1998kw}
\be
\phi=\tilde{\phi}-a\,x, \quad y \cos{\Theta}=\rho\cos{\theta}, \quad y^2= \frac{(\rho^2 \Delta_\theta+a^2 \sin^2{\theta})}{\Xi},
\ee
then at large $\rho$ it is equivalent to
\be
\d s^2 \approx -\frac{\d y^2}{y^2}+ \underbrace{y^2\Big(\d x^2 + \d \Theta^2+\sin^2{\Theta} \d\phi^2\Big)}_{=\gamma_{ij} \d x^i \d x^j}.
\ee
Indeed, the geometry contributes to the same wavefunction with spatial $S^1 \times S^2$, although the contractible circle requires imposing the identification \eqref{eq:idtwde}. 

\smallskip

We can evaluate the on-shell action directly in the cosmological geometry, or evaluating the Kerr-AdS action and performing carefully the analytic continuation. We obtain
\bea \label{eq:Kerronshell}
S_{\rm on-shell} &=& -\frac{ \L^2 L}{4G_N (1+a^2)}\left(\rho_+^3 + \rho_+(a^2+1)+\frac{ a^2}{\rho_+}\right) +   S_{\text{div}},\\ 
\label{eq:SdivKerr}
S_{\text{div}}&=& -\frac{\L^2}{4\pi G_N}\int_{S^1 \times S^2} \d^3 x\, \sqrt{\gamma} \left(1- \frac{1}{4} R[\gamma] \right).
\ea
More explicitly, the divergent term in the action is
\be
-\frac{\L^2}{4\pi G_N}\int_{S^1 \times S^2} \d^3 x\, \sqrt{\gamma} \left(1- \frac{1}{4} R[\gamma] \right)=-\frac{\L^2 L  }{6 G_N (1+a^2)}[6 \rho_b^3-\rho_b(3 +a^2-3a^3 \cos^2{\theta})].
\ee
Some relevant steps in the derivation of this expression are presented in Appendix \ref{app:ACTION}. The late-time geometry $\gamma$ represents a circle of radius $y \L$ and a sphere with radius $y$, at large times $y$. The term $S_{\text{div}}$ diverges in the late time limit. This term is purely imaginary since $\gamma$ is real, and does not affect the norm of the wavefunction. For that reason, we will ignore it in the discussion below although it gives an important contribution to the full wavefunction. 

\smallskip

Having described all the solutions, we need to decide which ones are physical and what their prediction for the wavefunction is. The equations for $\rho_+$ and $a$ have multiple solutions. Following the case with $\alpha=0$, we will look for those solutions that lie in the range $\text{Re}(\rho_+)>0$ and $\text{Im}(\rho_+)>0$ as well as those that satisfy $a\to 0$ as $\alpha \to 0$.

\smallskip

These conditions nail down a single solution for $\rho_+$ and $a$ for a given $L$ and $\alpha$. The explicitly formula is quite complicated and not very illuminating. Instead, we shall analyze it first in the Nariai limit $L\to\infty$. The physical solution satisfies
\bea
\rho_+ &=& \frac{1}{\sqrt{3}} + \frac{2\pi \i}{3 L} - \frac{4\pi^2 +3 \alpha^2}{6\sqrt{3} L^2} - \frac{\i \pi \alpha^2}{L^3} + \ldots ,\\
a&=& \frac{\alpha}{2L} + \frac{\i \sqrt{3} \pi \alpha}{L^2} - \frac{32\pi^2 \alpha + 3 \alpha^3}{8 L^3} + \ldots. 
\ea
The classical approximation to the wavefunction in the large $L$ limit, with fixed $\alpha$ is given by
\beq
\Psi[h] \sim \text{exp}\Big(- \frac{\i \mu_N S_{\text{dS}} L}{2 \pi} + \frac{S_{\text{dS}}}{3} + \i \frac{(8\pi^2 +\alpha^2)\mu_NS_{\text{dS}}}{8\pi L} - \frac{(16\pi^2 + 9 \alpha^2)S_{\text{dS}}}{54 L^2} + \ldots \Big).\label{eq:ONSAS1S2OMEGA}
\eeq
From this expression we can read off the coupling of the rotational $\SU(2)$ mode as the prefactor multiplying $\alpha^2$. Just like the Schwarzian mode, the coupling constant is imaginary and indeed we found in Section \ref{sec:rotaionalmodes} that the rotational eigenvalues tend to be purely imaginary as $L\to\infty$.

\smallskip

Let us now go back to our problem. Antiperiodic fermions along $S^1$ correspond to $\alpha=0$ while periodic fermions to $\alpha = 2\pi$. The explicitly answer for these two cases is
\bea
|\Psi(S^1_A \times S^2)|^2 &\sim& \text{exp}\Big( \frac{2 S_{\text{dS}}}{3} - \frac{16 \pi^2 S_{\text{dS}}}{27 L^2} + \ldots \Big),\\
|\Psi(S^1_P \times S^2)|^2 &\sim& \text{exp}\Big( \frac{2 S_{\text{dS}}}{3} - \frac{52 \pi^2 S_{\text{dS}}}{27 L^2} + \ldots \Big).
\ea

\smallskip

For a fixed $L$ the ratio between these partition functions is given by
\beq
\frac{|\Psi(S^1_A \times S^2)|^2}{|\Psi(S^1_P \times S^2)|^2} \sim e^{\frac{4 \pi^2 S_{\text{dS}}}{3 L^2}}\gg 1.
\eeq
For a given future spatial geometry, this suggests that it is non-perturbatively more likely to create a universe with antiperiodic fermions. Since the difference appears already at the level of the classical action we do not need to necessarily keep track of the quantum corrections.

\smallskip 
\begin{figure}
\centering   \includegraphics[width=0.6\linewidth]{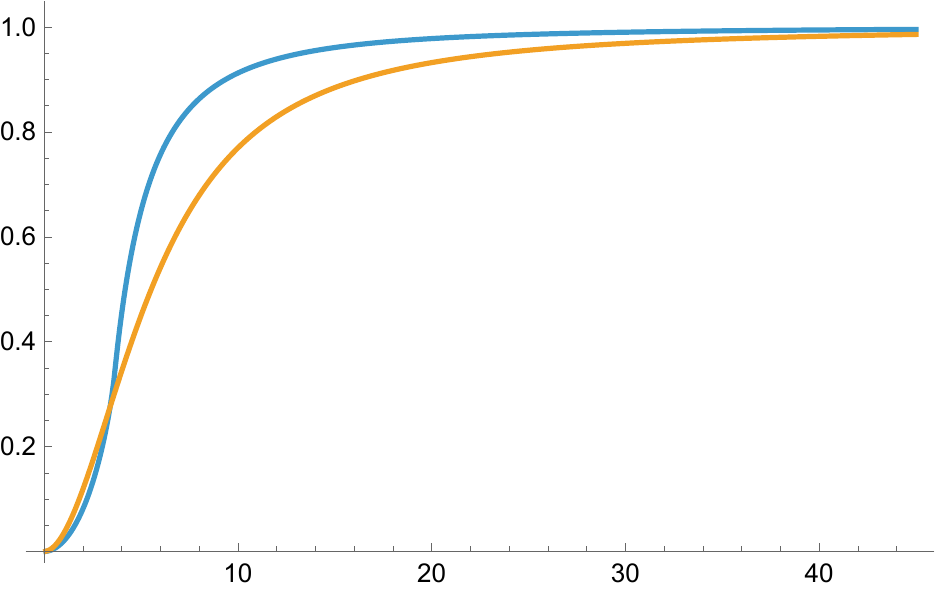}
\begin{picture}(0.3,0.4)(0,0)
\put(-270,175){\makebox(0,0){$\log |\Psi|^2$}}
\put(5,3){\makebox(0,0){$L$}}
\put(0,70){\makebox(0,0){\textcolor{blue}{\textbf{---}} Antiperiodic}}
\put(-10,60){\makebox(0,0){\textcolor{orange}{\textbf{---}} Periodic}}
\end{picture}
    \caption{Logarithm of the absolute value square of the wavefunction of an $S^1_L\times S^2$ universe, normalized by its value as $L\to\infty$. The blue (orange) line corresponds to a universe with fermions antiperiodic (periodic) along $S^1$. }
\label{fig:PeriodicVsAntiperiodic}
\end{figure}
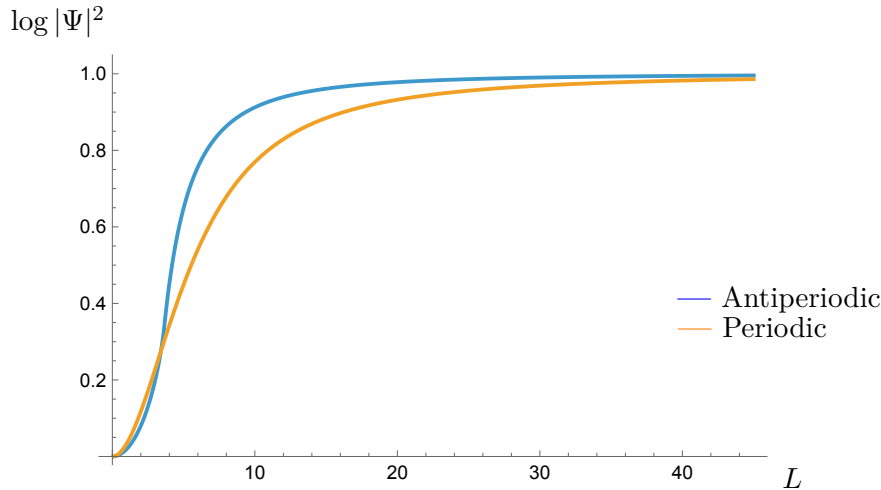

The conclusion of the previous paragraph is only valid if we compare them at the same $L$! When we maximize $|\Psi|^2$ over $L$ and compare, we get that to leading order, creating both spin structures are equally likely. The reason is that the change in $L$ for the two cases is such that the extremized norms in the classical approximation are equal. Concretely we get
\beq
L_P \sim \frac{2\pi \sqrt{26}}{3\sqrt{21}}\, \sqrt{S_{\text{dS}}},~~~~L_A \sim \frac{4\pi \sqrt{2}}{3\sqrt{21}}\,\sqrt{S_{\text{dS}}},
\eeq
where $L_A$ ($L_P$) is the most likely value for antiperiodic (periodic) fermions. The probability distribution for the most likely size of the universe in each case is given by
\beq
|\Psi(S^1_A \times S^2)|^2 \sim \frac{1}{L_A^{6}} \, e^{\frac{2 S_{\text{dS}}}{3}},~~~~~~|\Psi(S^1_P \times S^2)|^2 \sim \frac{1}{L_P^{6}} \, e^{\frac{2 S_{\text{dS}}}{3}},
\eeq
and therefore after incorporating quantum effects the probability of creating the two spin structures are of order one
\beq
\frac{|\Psi(S^1_P \times S^2)|^2}{|\Psi(S^1_A \times S^2)|^2} \sim O(1).
\eeq
This means that both universes are equally probable, since at most the ratio can be of order one, which is surprising. The remaining factors of $L$ in the one-loop determinants are slightly different $L_P/L_A \sim \sqrt{13/4}$, but not enough to produce an appreciable change.  It would be nonetheless interesting to determine that order one coefficient from a more detailed analysis of the one-loop determinants.

\section{Discussion}

To summarize, we have evaluated quantum corrections to the wavefunction of the universe for $S^1 \times S^2$ slices and verified that they render the wavefunction normalizable. We conclude here with some open questions we leave for future work.

\begin{itemize}  
\item We have found that creating a $S^1 \times S^2$ universe with periodic and antiperiodic fermions along $S^1$ is, to leading order, equally likely. It would be interesting to compute the one-loop determinants more carefully to precisely determine their relative probability. This would require evaluating the one-loop determinants in a geometry which is a section of a slowly rotating Kerr-dS black hole and therefore the eigenvalues can be computed in perturbation theory. It would be interesting to do for a slowly rotating charged near-extremal black hole as well. 
    \item Another open direction is to repeat the calculation done here to charged near-Nariai black holes. This is complicated by the coupling between the graviton and the photon. Some progress in this direction was done in \cite{Maulik:2025phe, Blacker:2025zca} by working in the near-horizon region, but one can adapt the methods introduced in \cite{Kolanowski:2024zrq} to extend their analysis to the full geometry.
    \item It was shown by Bousso and Hawking  \cite{Bousso:1997wi} that near-Nariai black holes can be perturbatively stable by incorporating matter one-loop effect \cite{Mukhanov:1994ax, Bousso:1997cg, Kummer:1998dc, Kummer:1999zy}. That is, there is an anti-evaporation channel that drives the near-Nariai black hole back to the degenerate Nariai limit, see also \cite{Nojiri:1998ue, Nojiri:1998ph} for a detailed analysis. The model under consideration contains a more realistic two-dimensional dilaton-coupled matter, dimensionally reduced from higher-dimensional minimally coupled theory.\footnote{Notably, a similar anti-evaporation instability was discovered with the same model for the Schwarzschild black hole \cite{Balbinot:1998yh, Buric:2000cj}, and it was considered unphysical due to its sensitivity to the choice of quantum state for the matter (see \cite{Wu:2023uyb} and references therein). However, in the near-Nariai regime, matter one-loop effects from the dilaton coupling are effectively subleading and the calculation is much less sensitive to the choice of state.} However, this is precisely the regime where quantum gravity corrections become important. It is therefore worth reconsidering the Hawking evaporation of a near-Nariai black hole, and incorporating the one-loop correction from the JT mode to the evaporation or anti-evaporation rate, similar to the work done recently for near-extremal Reissner-Nordstr\"om black holes \cite{Brown:2024ajk,Maulik:2025hax,Emparan:2025sao,Biggs:2025nzs}. This problem is complicated by the fact that in cosmology we do not have a spatial asymptotic region to anchor observers \cite{WOP_STW}.

  It was shown by Bousso and Hawking \cite{Bousso:1995cc, Bousso:1996au} that near-Nariai black holes naturally arise from pair creation during the vacuum-dominated inflation phase. The pair creation rate can be estimated by computing the no-boundary wavefunction with gravitational instantons corresponding to Nariai $S^2 \times S^2$ topology and the dS type $S^4$ topology \cite{Ginsparg:1982rs}. (Of course this interpretation is only reasonable if the partition function of $S^2 \times S^2$ is imaginary, but the overall phase of the gravitational path integral is subtle \cite{Polchinski:1988ua,Maldacena:2024spf,WOP_ST}.) Quantum fluctuations should then drive slightly away from the saddle point and result in a near-Nariai black hole.  The anti-evaporation mode can be excited from pair production, but is argued to be a transient effect that only happens initially based on the no-boundary condition \cite{Bousso:1997wi}; see however, a different perspective on this \cite{Nojiri:1998ph, Elizalde:1999dw}. It is viable that primordial black holes generated during inflation may have a much longer lifetime than expected.\footnote{We should still point out that if the interpretation about the partition functions is correct \cite{Ginsparg:1982rs} and there will be no Schwarzian corrections (but see also \cite{Bousso:1998na} for a different analysis on sub-Nariai pair production), the pair-created black holes are either exponentially suppressed in their production rate, or inflation will exponentially dilute their abundance, depending only weakly on the inflation models. Hence, it is highly unlikely for any astrophysical implications about dark matter, and any signals would be extremely rare from this mechanism. We thank Raphael Bousso for bringing up these important points to us.} With a full analysis from the quantum gravity corrections, we could ask about potential observational imprints if we have a strike of luck \cite{KM3NeT:2025npi, Boccia:2025hpm}.

\end{itemize}

\paragraph{Acknowledgements} It is a pleasure to thank Raphael Bousso, Alejandra Castro, Maciej Kolanowski, Juan Maldacena, Mukund Rangamani, Xiaoyi Shi, and Zhenbin Yang for useful discussions and previous collaborations on related topics. GJT and CHW are supported by the University of Washington and the DOE award DE-SC0011637. 

\begin{appendix}

\section{Evaluation of the on-shell action} \label{app:ACTION}

In this Appendix, we perform a direct evaluation of the on-shell actions of four-dimensional Lorentzian Schwarzschild-dS and Kerr-dS geometry that appear in Section \ref{sec:S1S2topology} and Section \ref{sec:wavefunctionspin}, respectively.

\smallskip

We consider the Einstein-Hilbert action with the appropriate Gibbons-Hawking boundary term
\be \label{eq:Loraction}
S=\frac{\L^2}{16 \pi G_N} \int \d^4 x \sqrt{-g} (R-6)+\frac{\epsilon\L^2}{8 \pi G_N} \oint \d^3x \sqrt{\gamma} K+\ldots,
\ee
where $\Lambda=3/\ell^2$ and we rescaled the metric by a factor of $\L^2$. Here $\epsilon=n^\mu n_\mu=-1$ with a spacelike cutoff boundary surface $\rho_b$ when the radii of both the $S^1$ and $S^2$ are large, but their ratio is fixed to be $L$, as explained in Section \ref{sec:S1S2topology}. It is not needed to include counterterms in the analysis, but we will still need to carefully evaluate the boundary term and extract the divergent part.

\smallskip

With the metric given in \eqref{eq:SdSmetric}, we have
\be
R=12, \quad \sqrt{-g}=\rho^2 \sin{\theta}, \quad L=\frac{ L_b \sqrt{f(\rho_b)}}{ \rho_b},
\ee
where the last relation is the ratio between the proper length of the $S^1$ circle at $\rho=\rho_b$ given by 
\be
L_{S^1}=\L \, \int_{0}^{L_b} \sqrt{|g_{xx}|}\, \d x=\ell L_b \sqrt{f(\rho_b)},
\ee
and the radius of $S^2$ at $\rho=\rho_b$. We evaluate first the bulk term of the on-shell action \eqref{eq:Loraction}
\bea
S_{\text{bulk}}&=&\frac{\L^2}{16 \pi G_N} 6 \int_0^{L_b} \d x \int_{\rho_+}^{\rho_b} \rho^2 \d \rho  \int \sin{\theta} \d \Omega_2 
\no\\
&=&\frac{L \L^2}{2 G_N} \frac{\rho_b}{\sqrt{f(\rho_b)}}(\rho^3_b-\rho^3_+)
\no\\
&\approx& \frac{L\L^2}{4  G_N} (-\mu-2 \rho^3_+) +\frac{L \L^2}{4 G_N }(2 \rho_b^3+ \rho_b)+\mathcal{O}\bigg(\frac{1}{\rho_b} \bigg),
\eea
where we have used the expansion of $f(\rho)$ at large $\rho$ with 
\be
\sqrt{f (\rho_b)} \approx \rho_b\bigg(1- \frac{1}{2\rho^2_b}+\frac{ \mu}{2 \rho^3_b}\bigg).
\ee
Note the terms that depend on $\rho_b$ will contribute to a divergent piece in the wavefunction. For the boundary term we pick the future-directed timelike normal $n^\mu= \sqrt{f(\rho)}\delta^\mu_\rho$ with $n^\mu n_\mu=-1$, then
\be
\sqrt{\gamma}=\rho^2 \sqrt{f(\rho)}\sin{\theta}, \quad K=\frac{2\sqrt{f(\rho)}}{\rho}+\frac{f'(\rho)}{2 \sqrt{f(\rho)}},
\ee
then
\bea
S_{\text{bdy}}&=&-\frac{\L^2}{8 \pi G_N}\int_0^{L_b}\d x\int \sin{\theta} \d \Omega^2 \bigg(2\rho f(\rho)+\frac{1}{2} \rho^2 f'(\rho) \bigg) \bigg|_{\rho=\rho_b}
\no\\
&\approx&-\frac{3  \L^2 L \rho_b^3}{2 G_N}+\frac{ L \ell^2 \rho_b}{4G_N}+ \mathcal{O}\bigg(\frac{1}{\rho_b} \bigg),
\eea
which at large $\rho_b$ will not generate any finite terms that do not depend on $\rho_b$. Hence the total on-shell action is
\bea
S_{\text{on-shell}}&=&S_{\text{bulk}}+S_{\text{bdy}}
\no\\
&=&\frac{L \L^2}{4  G_N} (- \mu-2 \rho^3_+)-\frac{L \L^2(2\rho^3_b- \rho_b)}{2 G_N},
\eea
where the second term is the $S_{\text{div}}$ in \eqref{eq:Sdiv1} that explicitly depends on the cutoff surface $\rho_b \to \infty$, and it receives contribution from both the bulk and boundary terms.
The on-shell action is used to evaluate the wavefunction $\Psi(S^1_L \times S^2) \propto e^{\i S_{\text{on-shell}}}$ given the choice of the physical saddle specified in Section \ref{sec:S1S2topology}. 

\smallskip

Here we also work out the on-shell action for the Kerr-dS geometry corresponding to the periodic fermion boundary condition in Section \ref{sec:wavefunctionspin}. With the metric given in \eqref{eq:KerrdS}, we work out 
\be
R=12, \quad \sqrt{-g}=\frac{\varrho^2 \sin{\theta}}{\Xi}, \quad  L \approx \frac{\ell L_b\sqrt{-\frac{\Delta_\rho (\rho_b)}{\varrho^2 (\rho_b)}}}{\rho_b}.
\ee
Then the bulk term of the on-shell action gives
\bea
S_{\text{bulk}}&=&\frac{\L^2}{16 \pi G_N} 6 \int_0^{ L_b} \d x  \int_0^{2 \pi} \d \phi\int_{\rho_+}^{\rho_b}  \int_0^\pi \frac{(\rho^2+a^2 \cos^2{\theta})\sin{\theta}}{\Xi} \d \theta \d \rho 
\no\\
&=&\frac{L\L^2}{2\Xi  G_N}\frac{\rho_b}{\sqrt{-\frac{\Delta_\rho(\rho_b)}{\varrho^2(\rho_b)}}}[(\rho_b^3-\rho_+^3)+a^2(\rho_b-\rho_+) ]
\no\\
&\approx&\frac{L \L^2}{4 \Xi  G_N} (- \mu-2 a^2 \rho_+-2 \rho_+^3)+
\no\\
&\quad&\frac{L \L^2}{4 \Xi  G_N } (2 \rho_b^3+\rho_b  +2\rho_b a^2 -\rho_b a^2\sin^2{\theta})+\mathcal{O}\bigg(\frac{1}{\rho_b} \bigg),
\eea
where we have approximated at large $\rho$
\be
\sqrt{-\frac{\Delta_\rho(\rho_b)}{\varrho^2(\rho_b)}} \approx \rho_b \bigg(1-\frac{(1-a^2 \sin^2{\theta})}{2\rho_b^2}+\frac{ \mu}{2 \rho_b^3} \bigg).
\ee
Similarly for the boundary term, we pick a future-directed timelike normal $n^\mu=\sqrt{-\Delta_\rho/\varrho^2} \delta^\mu_\rho$ with $n^\mu n_\mu=-1$ where
\be
\sqrt{\gamma}=\frac{\sqrt{-\Delta_\rho} \varrho \sin{\theta}}{\Xi}, \quad K=\frac{\sqrt{-\Delta_\rho}}{2 \varrho} \bigg(\frac{4 \rho}{\varrho^2}+\frac{\Delta_\rho'}{\Delta_\rho}-\frac{2\varrho' }{\rho} \bigg),
\ee
then the boundary term at large $\rho_b$ expansion
\be
S_{\text{bdy}}\approx -\frac{3  \L^2 L \rho^3_b}{2 (1+a^2)G_N}-\frac{ L \ell^2 \rho_b (a^2-9a^2 \cos^2{\theta}-3 )}{12 (1+a^2) G_N}+ O \bigg( \frac{1}{\rho_b} \bigg).
\ee
Again, no finite terms that do not depend on $\rho_b$ are generated. Hence the total on-shell action is
\bea
S_{\text{on-shell}}&=&\frac{L\L^2}{4 \Xi G_N} (- \mu-2 a^2 \rho_+-2 \rho_+^3)
\no\\
&\quad&-\frac{\L^2 L  }{6 G_N (1+a^2)}[6 \rho_b^3-\rho_b(3 +a^2-3a^3 \cos^2{\theta})],
\eea
where the second term corresponds to the $S_{\text{div}}$ in \eqref{eq:SdivKerr} that exactly agrees with a local integral on the boundary cutoff surface. By plugging in $\Xi$ and $\mu$ of the Kerr-dS geometry to the first finite term we recover \eqref{eq:Kerronshell}. As a consistency check, note that every expression above for the Kerr-dS geometry reduces to the previous on-shell calculation with Schwarzschild-dS geometry by setting $a \to 0$.

\section{Comments on the rotational modes}\label{app:SU2}

Consider the action of the rotational mode
\beq
\i S = K \int \d x\, \text{Tr}( g^{-1} \partial_x g)^2.
\eeq
The length of the $x$ circle is $L$ and we turn on a twist along this circle parametrized by $\alpha$, following the same conventions as in Section \ref{sec:wavefunctionspin}. The path integral of this theory is one-loop exact \cite{picken1989propagator} and given by
\beq\label{eq:appZrotc}
\Psi = \sum_{n\in\mathbb{Z}} \frac{K^{3/2}}{L^{3/2}} \frac{2(2\pi)^{1/2}(\alpha+4\pi n)}{\sin(\frac{\alpha}{2})} \, e^{- \frac{ K}{2L}(\alpha+4\pi n)^2}.
\eeq
The overall prefactor of $L^{-3/2}$ has the same origin as explained in Section \ref{sec:rotaionalmodes}. The sum over $n$ corresponds to a sum over saddles, as explained in Section 2 of \cite{Iliesiu:2021are} for example. 

\smallskip

In the case of the near-Nariai limit, we found that $K$ is purely imaginary with $K=-\i \mu_N S_{\text{dS}}/(4\pi)$. Although the sum over $n$ is clearly convergent whenever $\text{Re}(K)>0$, this property is lost in the Nariai wavefunction given that $\text{Re}(K)=0$. Naively, this would invalidate our analysis in Section \ref{sec:QuantumCorrections} and force us to consider non-perturbative corrections in the rotational mode sector. Fortunately, this is not so, but requires going away from the Nariai limit. We found in Section \ref{sec:wavefunctionspin} from \eqref{eq:ONSAS1S2OMEGA} that the leading behavior of the on-shell action is 
\beq \label{eq:ONSAS1S2OMEGA222}
\Psi[S_{L,\alpha}^1 \times S^2] \supset \text{exp}\Bigg( \underbrace{\i \frac{\alpha^2 \mu_N S_{\text{dS}}}{8\pi L} }_{\text{Reproduces action in \eqref{eq:appZrotc} }}- \underbrace{\frac{  9\alpha^2 S_{\text{dS}}}{54 L^2}}_{\text{Leading correction}} + \ldots \Bigg),
\eeq
where we only retain the $\alpha$-dependent terms. The non-perturbative configurations we need to sum over are obtained via the shift $\alpha \to \alpha + 4\pi n$. This is true in the Nariai limit and it is also true at the level of the 4d path integral where we should include shifts of the angular velocity, see for example Section 2 of \cite{Iliesiu:2021are}.  Therefore, even though the JT term, proportional to $1/L$, is imaginary and does not lead to a convergent sum over saddles, the leading correction is not only real, but has the appropriate sign to ensure that the sum over $n$ is convergent. This avoids any unphysical divergence in the path integral over the rotational modes.

\smallskip

Having determined that the higher-order terms in the large $L$ limit guarantee that \eqref{eq:appZrotc} is convergent, we can return to our analysis. In the analysis done in Section \ref{sec:QuantumCorrections}, we worked with $\alpha=0$. A careful analysis of this limit leads to
\beq
\Psi = \frac{K^{3/2}}{2\pi L^{3/2}} + \sum_{n\geq 1} e^{- \frac{8\pi^2 n^2 K}{L}}\left(\frac{K^{3/2}}{\pi L^{3/2}} - 16 \pi n^2 \frac{K^{5/2}}{L^{5/2}}\right).
\eeq
The answer is exact in two loops when $\alpha=0$, instead of one loop, since the space of fixed points is no longer isolated, as explained in Appendix A of \cite{Iliesiu:2021are} it is a two-dimensional manifold labeled by $n\geq 0$. In this expression, it is clear that terms with $n>0$ will be suppressed when $L\ll K \sim S_{\text{dS}}$ since the sum will be highly oscillatory. Therefore in this regime, we can approximate the rotational mode path integral by
\beq
\Psi \sim \frac{K^{3/2}}{2\pi L^{3/2}}.
\eeq
Up to an overall prefactor and phase, this matches the answer we found in Section \ref{sec:rotaionalmodes}. Moreover, it justifies that for the most likely size $L_0 \sim \sqrt{S_{\text{dS}}}$ we do not need to worry about non-perturbative corrections in this sector.


\end{appendix}

\bibliographystyle{utphys2}
{\bibliography{bibliography}{}}

\end{document}